\documentclass[apj]{emulateapj}
\usepackage{natbib}
\def\xmm{{\it XMM}}
\def\xmmnewton{{\it XMM-Newton}}

\defcitealias{FSN+10}{FSN}

\slugcomment{Submitted version of June 21, 2016}

\shorttitle{Merger Shock in Abell 3667}

\shortauthors{Sarazin et al.}

\begin{document}

\title{Deep {\it XMM-Newton} Observations of the NW Radio Relic Region of Abell 3667}

\author{%
Craig L. Sarazin\altaffilmark{1},
Alexis Finoguenov\altaffilmark{2,3},
Daniel R. Wik\altaffilmark{4},
and
Tracy E. Clarke\altaffilmark{5}
}

\altaffiltext{1}{%
Department of Astronomy, University of Virginia,
P. O. Box 400325, Charlottesville, VA 22904-4325, USA;
sarazin@virginia.edu}
\altaffiltext{2}{%
Department of Physics, University of Helsinki, Gustaf H\"allstr\"omin katu 2a,
FI-00014 Helsinki, Finland}
\altaffiltext{3}{%
Center for Space Science Technology, University of Maryland Baltimore
County, 1000 Hilltop Circle, Baltimore, MD 21250, USA}
\altaffiltext{4}{%
Astrophysics Science Division, 
NASA/Goddard Space Flight Center,
Greenbelt, MD 20771, USA; daniel.r.wik@nasa.gov}
\altaffiltext{5}{%
Naval Research Laboratory, 4555 Overlook Ave.\ SW,
Code 7213, Washington, DC 20375, USA; tracy.clarke@nrl.navy.mil}

\begin{abstract}
The results of long \xmmnewton\ X-ray observations of the NW radio relic of Abell 3667 are presented.
A shock is detected at the sharp outer edge of the radio relic,
both in the X-ray surface brightness and the temperature profiles.
The Mach number is
${\cal{M}} = 2.54^{+0.80}_{-0.43}$.
The temperature jump at the shock is larger than expected from the density jump, which may indicate that
a dynamically important magnetic field aligned primarily parallel to the shock front is present.
The gas temperature rises gradually
over several arc minutes within the shock region.
This could indicate that the shock energy is initially dissipated into some mix of thermal and nonthermal (e.g., turbulence) components, and that the nonthermal energy decays into
heat in the post-shock region.
The observed radio relic can be powered if $\sim$0.2\% of the energy
dissipated in the shock goes into the (re)acceleration of relativistic electrons.
We show that the observed steepening of the radio spectrum with distance behind
the shock is consistent with radiative losses by
the radio-emitting electrons.
However,  the radio spectrum immediately behind the shock is flatter
than expected for linear diffusive shock acceleration of thermal electrons.
This suggests that the shock
re-accelerates a pre-existing population of relativistic electrons.
We also detect a bright, cool region (the ``Mushroom'') to the south of the
radio relic, which we propose is the remnant cool core of a merging subcluster, and that this subcluster was the driver for the observed NW shock.
In this model, the properties of Abell 3667 are mainly the result of an offset binary merger, and the cluster is being observed about 1 Gyr after first core passage.
We predict that deeper X-ray or SZ observations of the SE radio relic will reveal a second merger shock at the outer edge.
\end{abstract}


\setcounter{footnote}{5}

\section{Introduction}
\label{sec:intro}

Clusters of galaxies are the largest relaxed systems in the Universe.
They form hierarchically by the merger of smaller systems.
While the most common mergers involve relatively small groups or
even individual galaxies,
major cluster mergers of similar-sized subclusters also occur with
a significant frequency.
These major cluster mergers are probably the most energetic events
which have occurred in the Universe since the Big Bang.
They drive shocks into the intracluster medium of the subclusters.
Although most merger shocks have relatively low Mach numbers of
${\cal{M}} \approx 1$--3,
shocks are the most important source of heating for the intracluster
medium in massive clusters.

Most collisionless astrophysical shocks in low density gas
also accelerate electrons (and presumably, ions) to relativistic energies.
In fact, extended diffuse cluster radio sources with steep spectra and no clear
optical counterparts have been known for over 40 years
\citep{Wil70, FGG+12}. 
They have such steep radio spectra that in most cases they
can be detected only at lower frequencies ($\la$5 GHz). 
Sources which are relatively symmetric and are projected on the cluster
center are often
referred to as ``radio halos'' \citep[e.g.,][]{DRL+97}, while
elongated sources usually located on the cluster periphery are called
``relics.''
In essentially every case, these diffuse cluster radio sources have
been found in irregular clusters that are undergoing mergers. 
This
suggests that the radio emitting electrons are accelerated or
re-accelerated by some merger-driven process.  One possible
theoretical picture is that halos are due to relativistic electrons
(re)accelerated by turbulence behind merger shocks, while the relics
are the direct result of merger shock acceleration
\citep[e.g.,][]{FGG+12}.
If the relics are due to shock
acceleration, then there should be a merger shock with its associated
temperature and density jump at one edge (generally the outer edge) of
the relic, and the relic radio spectrum should be flatter there due to
recent acceleration.  On the other hand, if a relic is due to
adiabatic compression of a pre-existing radio lobe, any merger shock
will likely be beyond the relic, and no clear radio spectral variation
is expected.

Abell~3667 is a violently merging, relatively low redshift
$(z=0.05525$) cluster \citep{MSV99}.  Early {\it Chandra} observations
detected a merger cold front near the center \citep{VMM01a,VMM01b},
and the term ``cold front'' was coined based in part on the Abell~3667 observations.
Subsequent {\it Chandra} and XMM/Newton observations \citep{MFV02,BFH04}
provided spectacularly detailed images and information on the dynamics
of the merger, but were mainly concentrated on the interior regions,
not the outer regions near the radio relics.

Abell 3667 contains a pair of curved cluster radio relics
\citep{RWH+97}, which are located on either side of the cluster center
at large radii.  Their locations and sharp, inwardly curved outer
edges are consistent with particle acceleration by merger shocks.  In
this picture, the shocks would be located at the outer edges of the
two relics.  The northwest radio relic in Abell 3667 is the brightest
cluster radio relic or halo source which is known, with
a flux at 20 cm of 3.7 Jy \citep{Joh-Hol04}.  It is also one of the
largest relics, with a total extent of 33\arcmin\ or 2.1 Mpc.  Its
radio spectrum steepens with distance from the outer edge
\citep{RWH+97} as expected if the electrons are accelerated there and
the higher energy electrons lose energy due to synchrotron and IC
emission as they are advected away from the shock.

\citet{RBS99} presented a numerical MHD/N-body simulation of Abell 3667 with parameters
chosen to match crudely the observed properties of the cluster.
In their model, Abell 3667 is an offset binary merger which is
currently $\sim$1 Gyr beyond first core passage.
The mass ratio in their model was 5:1.
The smaller subcluster merged from the SE, and has driven a shock ahead of
it which corresponds in position to the NW radio relic.
A second shock was driven by the primary cluster, and is
at the location of the SE radio relic.
There were also a few weaker and smaller shocks near the center
of the cluster without clear radio counterparts.
More recently, \citet{DSB+14} presented a more complicated
MHD/N-body simulation with initial conditions drawn from a cosmological
simulation.
This simulation was not tuned to represent Abell 3667 specifically,
but does show many of the same features as the \citet{RBS99} simulation.

Recently, diffuse radio emission has been detected near the center of Abell 3667.
Parkes and Australia Telescope Compact Array observations show an unpolarized
radio ``bridge'' running from the NW radio relic to the center of the cluster
\citep{Car+13}.
They argue that this bridge is a turbulent wake associated with a merging subcluster
and the merger shock which made the NW radio relic.
Observations with the Karoo Array Telescope (KAT-7) don't detect the bridge,
but do find a possible radio mini-halo associated with the brightest cluster galaxy,
B2007$-$569
\citep{RSO+15}.

Here, we present the results of a series of long {\it XMM-Newton}
observations of Abell~3667 which cover the region of the NW radio
relic.
The X-ray observations and data analysis are described in \S~2.
The X-ray images and surface brightness profiles are given in
\S~3.
In \S~4, the X-ray spectra and cluster temperature map are presented.
We discuss the implications of these observations in \S~5.
Our conclusions are summarized in \S~6.
Appendices give the expressions used to model shocks observed in the
X-ray image.
We assume a $\Lambda$CDM cosmology with $H_0 = 71$ km s$^{-1}$
Mpc$^{-1}$, $\Omega_M=0.27$, and $\Omega_\lambda=0.73$.  At the
redshift of Abell 3667 of $z=0.05525$, the angular diameter distance
is $d_A = 212.6$ Mpc, and 1\arcmin\ corresponds to 61.8 kpc.  For a
mean cluster temperature of 7.2 keV, the virial radius should be
$r_{180} \approx 2.33$ Mpc $= 38\arcmin$ \citep{MFS+98}.
(Within the radius $r_{180}$, the average density is 180 times
the cosmological critical density at the cluster redshift.)
Unless otherwise specified, we provide confidence intervals at the 68\%
level.

\begin{figure}[t]
 \begin{center}
 \includegraphics[scale=0.35, angle=0]{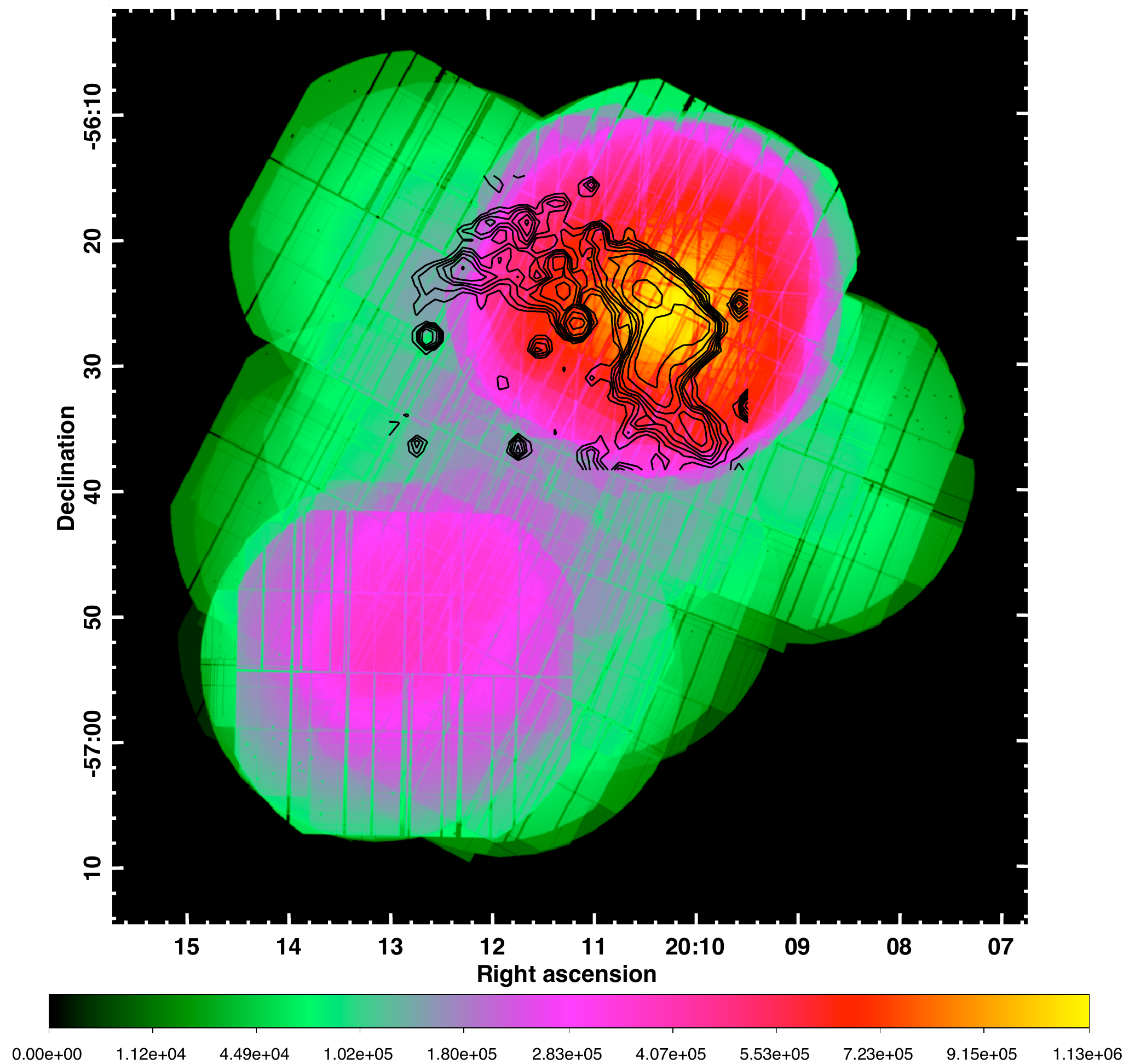}
 \end{center}
 \vskip-0.1in
 \caption{The exposure map showing the total amount of good time for all of the images and all of the observations
 of Abell 3667.
 The color bar at the bottom shows the total exposure in seconds with a square-root scaling.
 The SUMSS 843 MHz radio contours from the region of the NW radio relic are shown 
 black (8 countours logarithmically distributed between 1 and 23 mJy beam$^{-1}$).
 }
 \label{fig:expmap}
\end{figure}

\section{X-ray Observation and Data Reduction}
\label{sec:obs}

The eight previous {\it XMM-Newton} observations of Abell 3667 were
discussed in
\citet[][hereafter \citetalias{FSN+10}]{FSN+10};
six of these
observations, which concentrate on the center of the cluster, were
analyzed in \citet{BFH04}.  We observed the northwest region of the
cluster, including the area of the NW radio relic, with {\it
  XMM-Newton} in orbit 1620 (OBSID 0553180101).  Based on the results
from this observation, we proposed a much longer observation of the NW
region of the cluster, which was done on 2010 September 21--27,
October 3, November 2--3.  There were four pointings centered on the
NW radio relic (OBSIDs 065305401, 065305501, 065305601, and 065305601
in orbits 1976--1978 and 1996) for a total of 290 ksec, and two
pointings on either side (east and west) of the relic (OBSIDs
065305201 and 065305301 in orbits 1975 and 1981) of 30.0 and 26.0 ksec.
The medium filter has been used for all instruments for consistency
with the previous observations of the cluster.
The observations were done in extended full frame
mode for the pn to reduce the effect of out-of-time
events.
We checked whether any of the CCDs were in anomalous states
with unusually high background, and removed CCD4 of MOS1 from observations 0553180101, 0653050301, 0653050401, 0653050501 and CCD5 of MOS2 from 0553180101.

\begin{figure*}[t]
 \begin{center}
 \includegraphics[scale=0.35, angle=0]{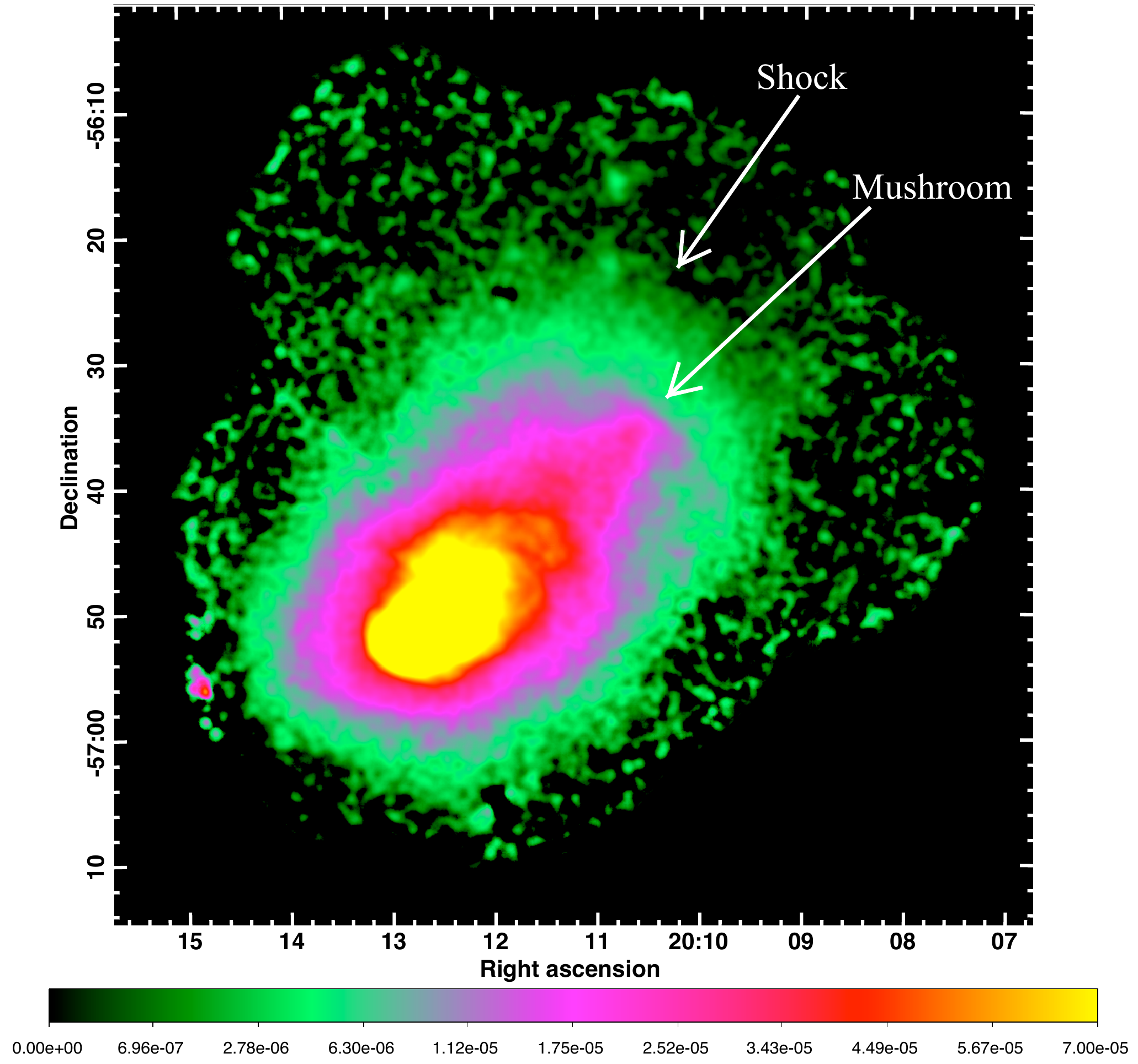}
\includegraphics[scale=0.47, angle=0, trim=0 10 0 0]{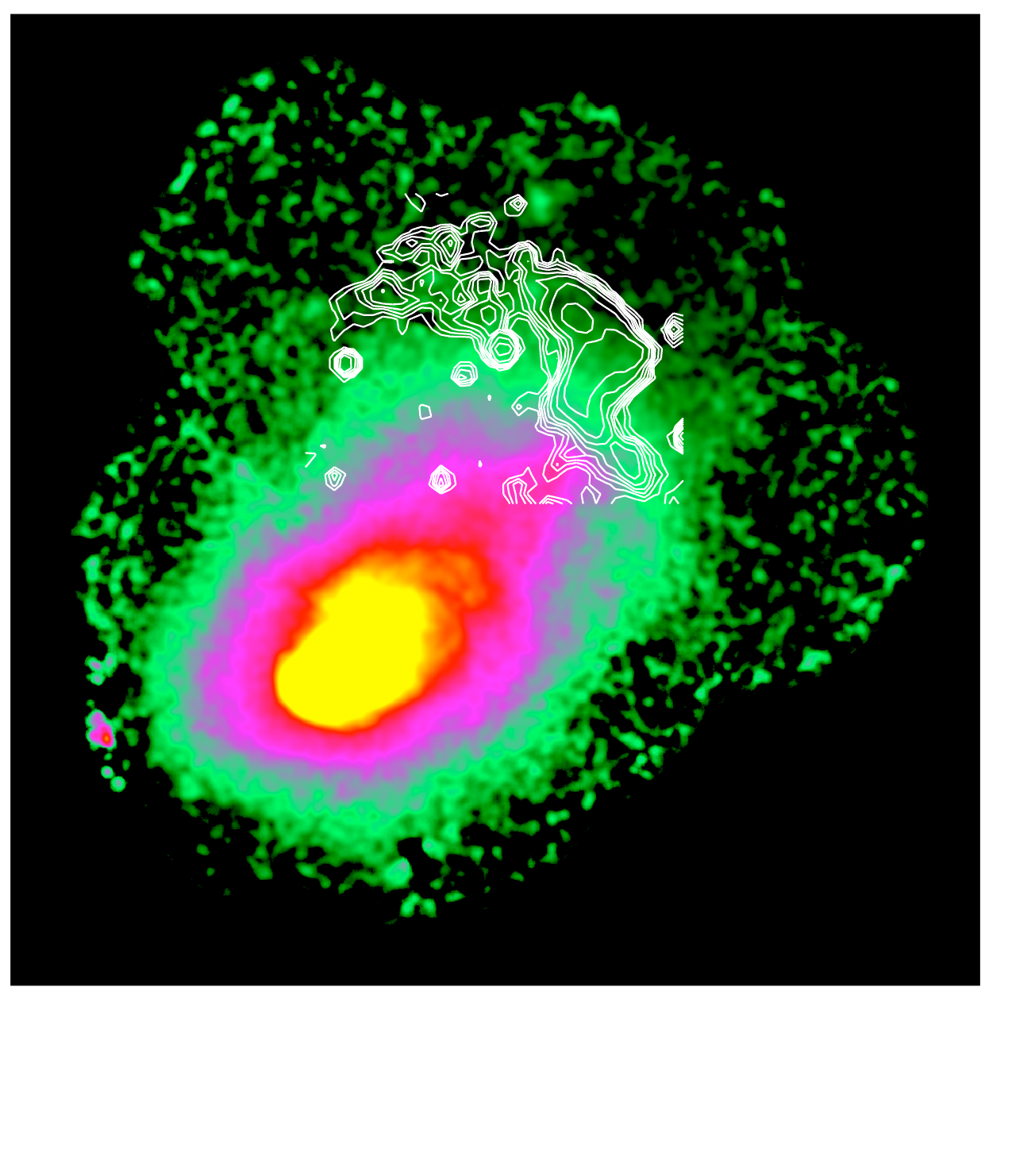}
 \end{center}
 \caption{X-ray mosaic image from {\it XMM-Newton} in the 0.5--2.0 keV band (excluding the instrumental Al line at $\sim$1.5 keV).
 the image has been corrected for exposure and background, and all point sources have been removed from the image.
 {\it Left:} {\it XMM} image with NW features labelled.  {\it Right:} {\it XMM} image with SUMSS 843 MHz radio contours shown.
 }
 \label{fig:whole_image}
\end{figure*}

We used {\sc sas} version 13.5.0%
\footnote{http://xmm.esac.esa.int/sas/current/howtousesas.shtml} for
the standard data reduction.  After producing the event files, we have
applied a strict light curve cleaning
\citep[e.g.,][]{ZFB+04}
in
order to ensure that we retained only the periods with the lowest and
the most stable background.  The new data was combined with all of the
previous observations to make a mosaic.  Figure~\ref{fig:expmap} gives
the total exposure maps showing the total of all of the good time for
all of the observations and for the sum of MOS1, MOS2, and pn
detectors.  The maximum total exposure, in the center of the NW Radio
Relic, is $1.13 \times 10^6$ sec.

The quadruple background subtraction technique of \citet{Fin+07} was
used to model the background, and remove it from the images and
surface brightness profiles.

The point sources in the X-ray mosaic were detected, modeled using the
{\it XMM}
PSF, and subtracted from the images and surface brightness profiles
described below.
For the spectral analysis, we excised the regions where
contribution from point source emission could be significant, and in a
similar way we excised the detected point sources in the background
file, so remaining emission from the wings of PSF which is confused,
is subtracted by using the background file.

The instrumental background level at the epoch of latest Abell~3667
observations was substantially higher than the background
level typical for the publicly available calibration products and
the quality of the background subtraction was not satisfactory.
To
remedy the problem, we have used the {\it XMM} observations of the
{\it Chandra} Deep Field South (CDFS) field
\citep{Com+11,Fin+15},
which were performed at a similar epoch to A3667 observations.
We have
chosen with work with the pn detector, for which we have
confidence in our knowledge of the instrumental characteristics. We
collected 265 ksec of cleaned data of the CDFS observations performed in
2010 and masked out the areas where strong point sources were present.
The different choice of filters for the CDFS observations precludes using energies below 0.8
keV in the spectral analysis, so we have concentrated on the 0.8-7.5
keV band, but excluded the energies around the instrumental Al line at 1.5 keV. 

Spectral fitting was done using version 12.6.0 of {\sc xspec}%
\footnote{See
\url{http://heasarc.gsfc.nasa.gov/docs/software/lheasoft/}\label{ftn:heasoft}.}.
In addition to a custom made background subtraction, we
added the background power law component (the component that is not
multiplied with the effective area of the telescope) to fit for
possible soft protons present in the observations. We fixed the slope
of the components using the outmost spectral extraction region and
allowed the normalization to vary.

\section{X-ray Images}
\label{sec:images}

The {\it XMM-Newton} X-ray mosaic image in the 0.5--2.0 keV band (excluding the instrumental Al line at $\sim$1.5 keV) is shown in
Figure~\ref{fig:whole_image}
after removing the point sources.
This image was smoothed with a 36\arcsec\ gaussian to better bring out the fainter features in the NW region.
Here, we concentrate on two features to the NW of the cluster center, which were also discussed in
\citetalias{FSN+10}.
First, there is an elongated filament of brighter X-ray emission, which ends in a flattened top, which we label the ``Mushroom''.
There is a bright concentration of X-ray emission just below the top of the Mushroom, which coincides with the center of one of the major
subclusters of galaxies within Abell~3667 \citep[KMM2; see][]{OCN09}.
The top of the Mushroom coincides with the sharp SW bottom edge of the NW radio relic.

Second, there is a sharp drop in the diffuse X-ray surface brightness further to the NW.  This drop, which is the main subject of the current paper, is discussed in more detail below
(\S\S~\ref{sec:sb_shock}, \ref{sec:spectra_tmap}).
It coincides with the sharp outer edge of the NW radio relic.

\subsection{X-ray Surface Brightness Profiles}
\label{sec:sb}

The radial X-ray surface brightness (SB) profile from the center of Abell~3667 to the NW was determined in the 0.5--2.0 keV band (excluding the Al line at $\sim$1.5 keV).
Point sources were modeled using the  \xmmnewton\ PSF, and their flux was removed from the surface brightness profile.
The surface brightness was corrected for exposure and background.
The shapes of the elliptical pie annuli used to accumulate the SB profile are shown in Figure~\ref{fig:sb_regions} below.
Figure~\ref{fig:sb_cluster} shows the radial surface brightness profile of the cluster along the NW direction.
There are two down-turns (``knees'') in the outer part of the profile which correspond to the locations of the Mushroom and the Shock features.
The radial region occupied by the NW radio relic lies between these two X-ray features.

\begin{figure}[t]
 \begin{center}
 \includegraphics[scale=0.55, angle=0]{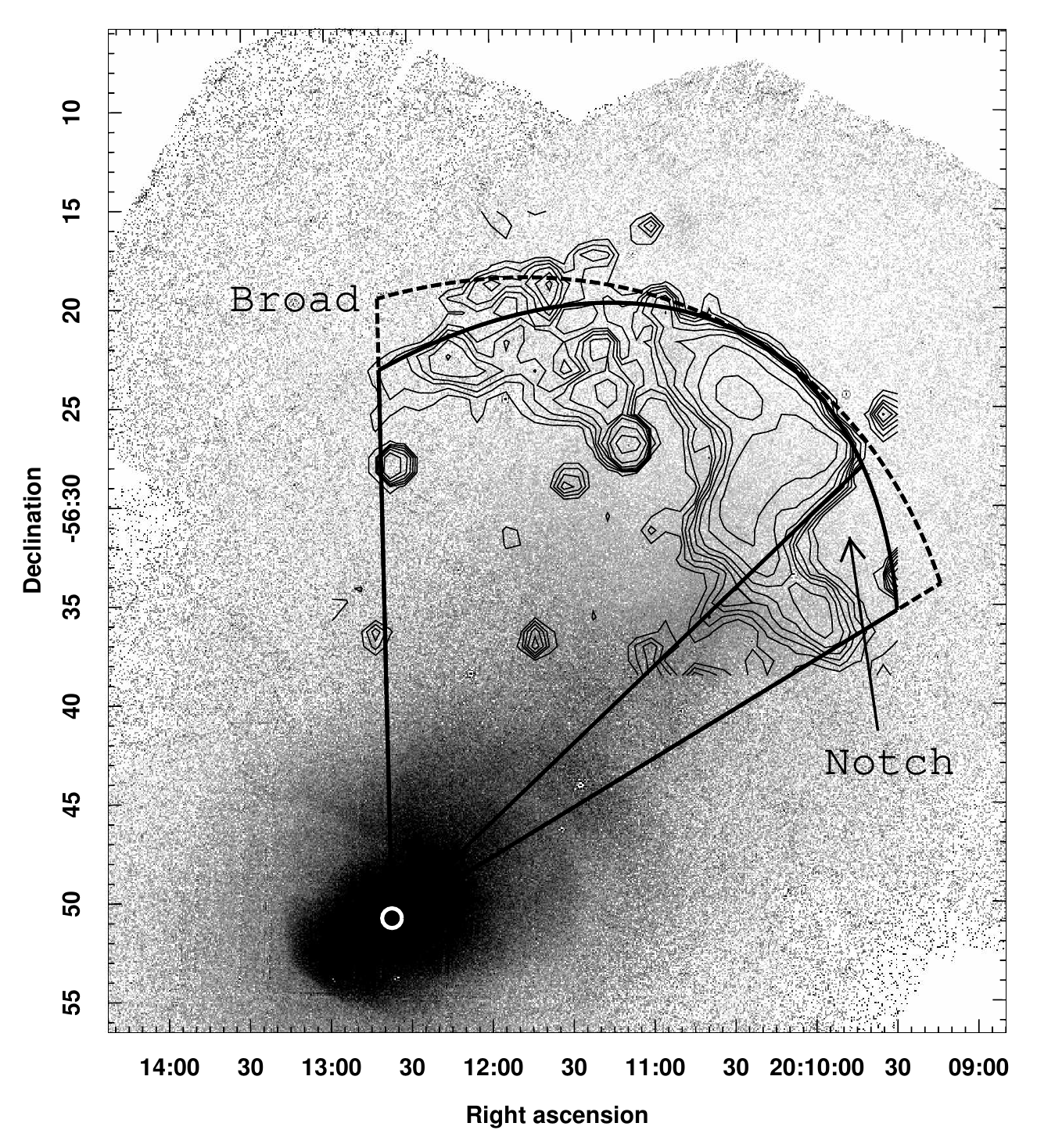}
 \end{center}
 \vskip-0.1in
 \caption{Shapes of the elliptical annular wedges used to derive the radial X-ray surface brightness profile across the radio relic are shown.
 The greyscale image is the \xmmnewton\ X-ray image with the point sources removed.
 The contours are the SUMMS 843 MHz radio image.
 The solid lines show the shape of the default elliptical wedges used to derive the surface brightnesses;
 the wider region includes the ``notch'' in the radio relic, while the narrower region excludes it.
 This wider region was also used to accumulate the full cluster profile in the NW regions
 (\S~\ref{sec:sb}).
 The dashed curve shows the ``broad'' region for the relic, including the radio extension to the NE.
 In each case, the elliptical wedge shown corresponds to our adopted shape for the outer edge of the radio relic.
 The regions all start at the cluster pressure peak, which is indicated by the white circle.
 }
 \label{fig:sb_regions}
\end{figure}

\subsection{X-ray Surface Brightness Profile across the Relic}
\label{sec:sb_shock}

The radial X-ray surface brightness profile across the NW radio relic was determined in the 0.5--2.0 keV band (excluding the Al line at $\sim$1.5 keV).
The surface brightness was corrected for exposure and background.
We assumed that the X-ray surface brightness is constant on self-similar ellipses.
The shape of the ellipses was taken to match approximately the sharp outer edge of the radio relic.
The surface brightness was accumulated  in elliptical pie annuli (``epanda'' regions).
In Figure~\ref{fig:sb_regions},
the ellipses shown correspond to our estimates of the outer edge of the relic.

\begin{figure}
 \begin{center}
\includegraphics[scale=0.43, angle=0]{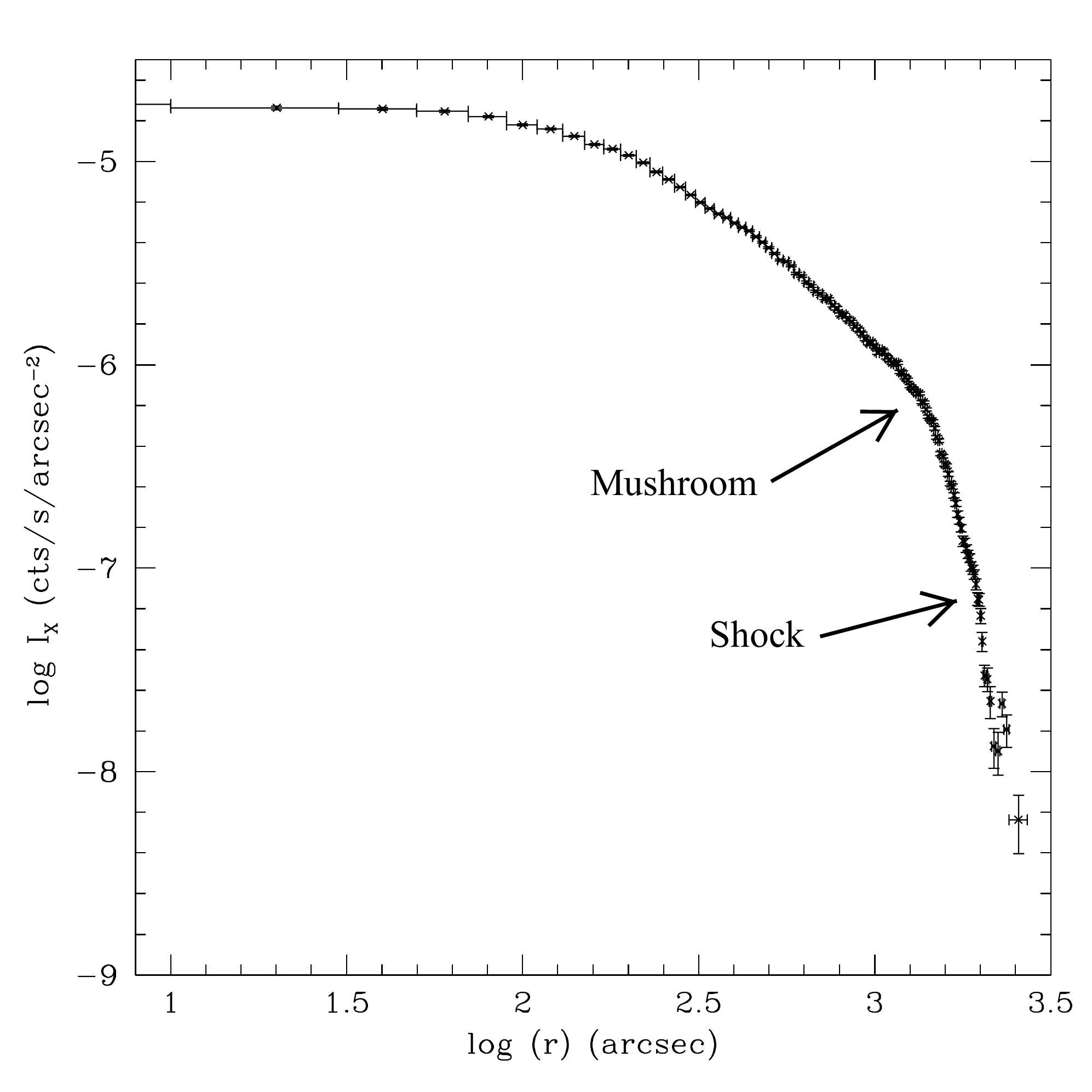}
   \end{center}
   \vskip-0.2in
 \caption{Radial X-ray surface brightness profile of Abell~3667 from the cluster center to beyond the NW radio relic.
 The profile has been corrected for exposure and background, and point sources have been removed.
 The approximate positions of the top of the Mushroom feature and the Shock are indicated.
 The NW radio relic lies roughly between these two features.
 }
\label{fig:sb_cluster}
\end{figure}

For all these regions, the major axis of the ellipse had a position angle of $-$33$^\circ$ measured from north to east.
As a default, we considered a wedge whose width was taken to be approximately the full width of the relic.
This wedge extends from an angle of $-$26$^\circ$ to $+$34$^\circ$ of the semi-major axis (measured counter-clockwise).

\begin{figure}[t]
 \begin{center}
 \includegraphics[scale=0.43, angle=0]{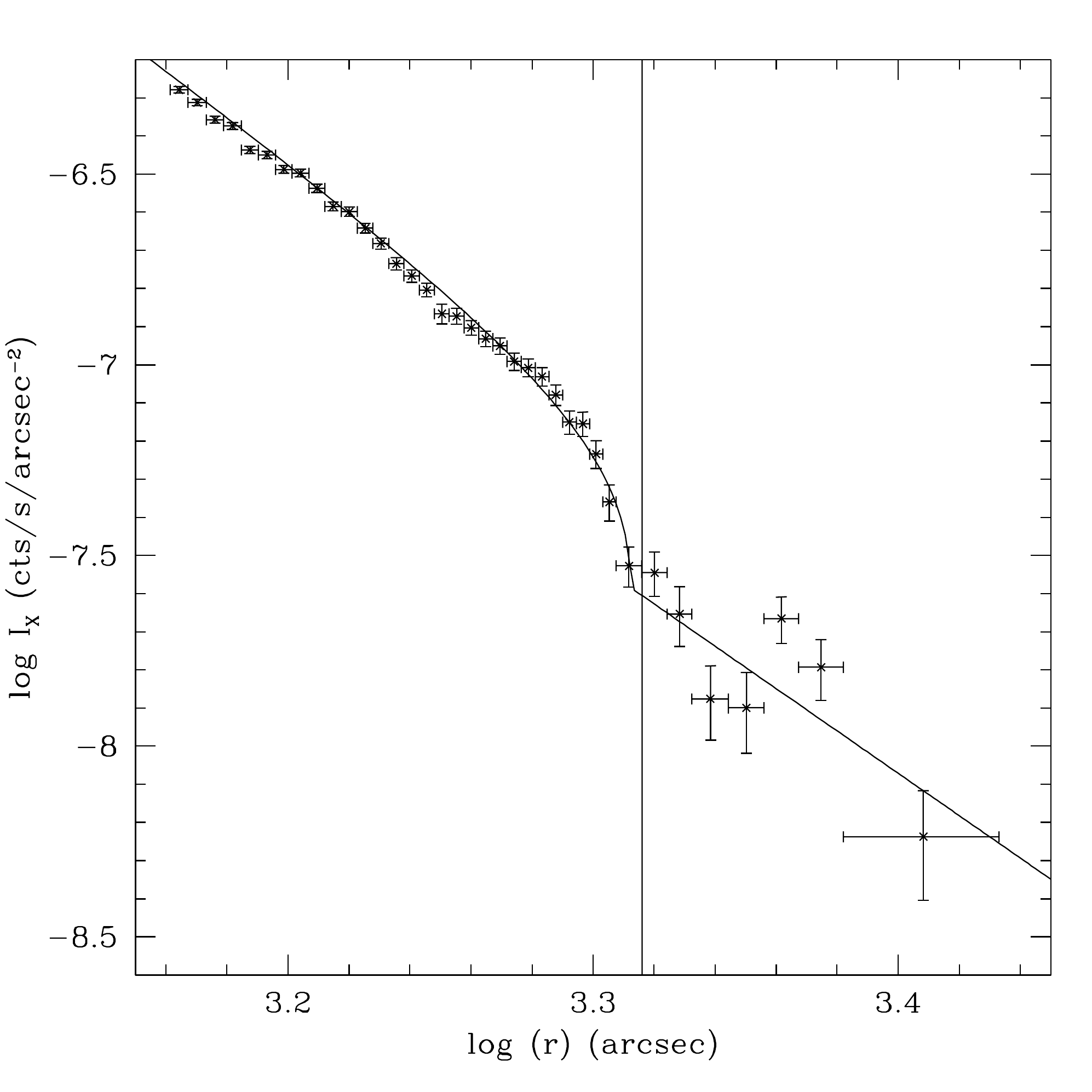}
 \end{center}
 \vskip-0.2in
 \caption{Radial X-ray surface brightness profile across the radio relic in the 0.5--2.0 keV band, excluding the Al line at $\sim$1.5 keV.
 The profile was corrected for exposure and background, and point sources were removed.
 The crosses give the data points, with 1-sigma uncertainties and the widths of the bins shown.
 An emissivity jump is evident very near to the outer edge of the relic ($r_{\rm relic}$), which we interpret as a shock.
 The solid curve is the best-fit surface brightness model for an ellipsoidal emissivity edge.
 The vertical line is the estimated position of the outer edge of the radio relic ($r_{\rm relic} = 2070\arcsec$), which is just outside of (but consistent within the errors with) the X-ray  edge.
 }
 \label{fig:sb_shock}
\end{figure}

The values of the surface brightness are shown in Figure~\ref{fig:sb_shock}.
The radii $r$ and $r_{\rm relic}$ are measured along the semi-major axis.
In the inner part of the profile, the bins had a width of 20\arcsec, and they increased in the outer parts to maintain a reasonable signal to noise ratio.
Note that the uncertainties increase dramatically for the outer points where the surface brightness is quite low.
The surface brightness profile is concave downward and then changes slope at a radius which is close to the outer edge of the relic.

We initially attempted to fit the profile with a standard ``beta-model''.
This gave a poor fit, with a $\chi^2$/dof = 5.16.
The value of the core radius was very small and undetermined, since the profile here only included the outer parts of the cluster.
The value of $\beta$ was remarkably large, $\beta = 2.63 \pm 0.06$.
This implies a very steep fall-off in the gas density in these regions.
With the small value of the core radius, the beta-model reduces to a power-law.
The X-ray surface brightness for the beta-model at large radii varies as $I_X \propto r^{-6 \beta + 1}$, which gives a variation of roughly $I_X \sim r^{-15}$.
Because uncertainties on the surface brightness grow rapidly with radius, the best-fit beta-model is essentially a power-law which fits the inner points of the profile in Figure~\ref{fig:sb_shock}, but lies above all of the data points are large radii.
Thus, the beta-model fit misses completely the concave downward drop and change in slope of the observed surface brightness.

To fit this discontinuity, we assumed that the X-ray emissivity is constant on self-similar ellipsoidal surfaces, and that a principal axis of these ellipsoids is in the plane of the sky.
The ellipsoidal surfaces are assumed to be symmetric about this principle axis, so that the surfaces of constant emissivity are actually prolate or oblate spheroids, and the extent along the line-of-sight is that same as the projected extent on the sky.
The elliptical shapes of the projection of the ellipsoids on the sky was taken to be similar to the outer edge of the relic
(the ellipses in Fig.~\ref{fig:sb_regions}).
This simple, self-similar ellipsoidal model represents a significant source of systematic errors in our analysis.
Some of these assumptions (the extent along the line-of-sight) affect the the overall scale of the emission (e.g., the
gas densities may all be off by a constant factor), but some of the other assumptions may affect the shock jump conditions we derive.

We assume that the emissivity had an ellipsoidal discontinuity, but was a power-law function of the elliptical radius inside and outside of this discontinuity.
If $r_{\rm edge}$ is the radius of this discontinuity, then the X-ray emissivity was assumed to vary as
\begin{equation}
\epsilon ( r ) = \left\{
\begin{array}{ll}
\epsilon_i ( r / r_{\rm edge} )^{-2 p_i } & r < r_{\rm edge} \\
\epsilon_o ( r / r_{\rm edge} )^{-2 p_o } & r > r_{\rm edge} \, .
\end{array}
\right.
\label{eq:emiss}
\end{equation}
The X-ray surface brightness is then given by
\begin{equation}
I_X (r) = I_{\rm in} (r) + I_{\rm out} (r) \, ,
\label{eq:sb}
\end{equation}
where
\begin{equation}
I_{\rm in} (r) = I_i A^{-2 p_i +1}  \left\{
\begin{array}{ll}
1 - I_{A^2} ( p_i - \frac{1}{2} , \frac{1}{2} ) & A^2 < 1 \\
0                                          & A^2 \ge 1 \, ,
\end{array}
\right.
\label{eq:sbi}
\end{equation}
and
\begin{equation}
I_{\rm out} (r) = I_o A^{-2 p_o +1}  \left\{
\begin{array}{ll}
I_{A^2} ( p_o - \frac{1}{2} , \frac{1}{2} ) & A^2 < 1 \\
1                                      & A^2 \ge 1 \, .
\end{array}
\right.
\label{eq:sbo}
\end{equation}
Here, $A \equiv ( r / r_{\rm edge})$,
$ I_{x} ( a , b ) \equiv B_x ( a , b ) / B ( a, b)$ is the normalized incomplete beta function,
$B_x ( a , b ) \equiv \int_0^x t^{a-1} (1-t)^{b-1} \, dt$ is the incomplete beta function,
$ B(a,b) = \Gamma (a) \Gamma (b) / \Gamma (a+b)$ is the beta function, and $\Gamma(a)$ is the gamma function.
For details and the expressions for accumulating counts in bins, see Appendix~\ref{sec:ellipt} and also
\citet{VMM01b}
and
\citet{Kor+11}.
The value $I_o$ gives the surface brightness at the edge, while
\begin{equation}
I_i = R \, I_o \frac{ B( p_i - \frac{1}{2} , \frac{1}{2} )}{ B( p_o - \frac{1}{2} , \frac{1}{2} )}
\label{eq:ii}
\end{equation}
normalizes the surface brightness due to gas within the edge.
Here, $R \equiv \epsilon_i / \epsilon_o $ gives the jump in X-ray emissivity at the edge.


\begin{deluxetable*}{lccccccc}
\tablewidth{0pt}
\tabletypesize{\scriptsize}
\tablecaption{Radio Relic and Mushroom X-ray Surface Brightness Profile Fits\tablenotemark{a}
\label{tab:shock_sb}}
\tablehead{
&
\colhead{$r_{\rm edge}$} &
\colhead{$\Delta r_{\rm edge}$} &
&
&
&
\colhead{$I_o$}
&
\\
\colhead{Region} &
\colhead{(arcsec)} &
\colhead{(arcsec)} &
\colhead{$p_i$} &
\colhead{$p_o$} &
\colhead{Ratio $R$} &
\colhead{($10^{-8}$ cts s$^{-1}$ arcsec$^{-2}$)} &
\colhead{$\chi^2$/dof}
}
\startdata
Relic with notch & $2053^{+17}_{-20}$ & $-17^{+17}_{-20}$ & $+3.40^{+0.09}_{-0.10}$ & $3.30^{+1.10}_{-1.02}$ & $3.11^{+1.75}_{-0.94}$ &
\phn$2.61^{+0.67}_{-0.53}$ & \phn51.94/32 $=$ 1.654 \\[4pt]
                          & (2070)                      & (0)                           & $+3.43^{+0.09}_{-0.09}$ & $2.88^{+1.06}_{-0.94}$ & $3.74^{+2.20}_{-1.13}$ &
\phn$2.22^{+0.41}_{-0.39}$ & \phn54.60/33 $=$ 1.894 \\ [4pt]
                           & (2070)                     & (0)                           & $+3.57^{+0.25}_{-0.25}$ & $8.68^{+5.13}_{-2.41}$ & (1) &
\phn$4.04^{+0.53}_{-0.66}$ & 102.60/34 $=$ 3.018 \\[4pt]                        
Relic without notch & $2059^{+28}_{-23}$ & $-11^{+28}_{-23}$ & $+3.39^{+0.13}_{-0.14}$ & $3.56^{+2.43}_{-2.55}$ & $4.73^{+18.71}_{-2.52}$ &
\phn$1.02^{+0.62}_{-0.45}$ & \phn59.89/32 $=$ 1.871 \\[4pt]
                           & (2070)                     & (0)                               & $+3.41^{+0.11}_{-0.12}$ & $2.86^{+2.86}_{-2.31}$ & $6.43^{+96.77}_{-2.96}$ &
\phn$0.84^{+0.39}_{-0.34}$ & \phn60.73/33 $=$ 1.840 \\[4pt]
Relic, broad region & $2034^{+30}_{-27}$ & $-36^{+30}_{-27}$ & $+3.69^{+0.13}_{-0.12}$ & $3.29^{+1.57}_{-1.73}$ & $2.90^{+3.81}_{-1.04}$ &
\phn$2.22^{+0.76}_{-0.67}$ & \phn76.85/32 $=$ 2.402  \\[4pt]
                            & (2070)                   & (0)                                 & $+3.76^{+0.10}_{-0.11}$ & \phn$2.43^{+1.78}_{-1.69}$ & $4.24^{+17.18}_{-1.86}$ &
\phn$1.66^{+0.47}_{-0.44}$ & \phn80.92/33 $=$ 2.452  \\[4pt]
Mushroom           & $1503^{+4}_{-5}$ &                                       & $-0.13^{+0.08}_{-0.09}$ & \phn$3.06^{+0.22}_{-0.20}$ & $2.75^{+0.25}_{-0.25}$ &
$44.3^{+2.2} _{-1.8}$    & \phn55.38/36 $=$ 1.538  \\[4pt]
                            & (1503)                  &                                       & $+0.53^{+0.06}_{-0.06}$ & \phn$5.43^{+0.12}_{-0.12}$ & (1) &
$65.8^{+0.7}_{-0.7}$     & 698.66/38 $=$ 18.396  \\[4pt]
\enddata
\tablenotetext{a}{Values in parentheses were held fixed during the fit.}
\end{deluxetable*}

This surface brightness model was convolved with the PSF of \xmmnewton\
\citep{RRS+11}%
\footnote{http://xmm2.esac.esa.int/docs/documents/CAL-TN-0029-1-0.ps.gz},
and then
accumulated in the bins used to determine the observed surface brightness
(see Appendix~\ref{sec:ellipt} for details).
The variable parameters of the model were taken to be the radius of the edge ($r_{\rm edge}$), the emissivity power-law exponents within and outside of the edge ($p_i$ and $p_o$, respectively), the emissivity jump at the edge $R$, and the surface brightness at the edge ($I_o$).
Because the value of $r_{\rm edge}$ is measured from the somewhat arbitrary center of the ellipse used to fit the outer edge of the radio relic, we instead report the value of $\Delta r_{\rm edge} \equiv r_{\rm edge} - r_{\rm relic}$, where
$r_{\rm relic}$ is the radius of the outer edge of the radio relic as shown in Figure~\ref{fig:sb_regions}.
The model parameters were varied until the value of $\chi^2$ was minimized.
The uncertainty (90\% confidence) in each of the parameters was determined by varying all of the variables, and determining the width of the surface where the value of $\chi^2$ was increased by 2.706, marginalizing over all the other parameters.
The best-fit values of the parameters, the minimum value of $\chi^2$,
and the number of degrees of freedom (dof)
are all given in Table~\ref{tab:shock_sb}.

We first considered regions which were essentially the full width of the radio relic, and which fit well the sharp outer edge at the center of the relic (the wider solid lines in Fig.~\ref{fig:sb_regions}).
The fits to this surface brightness profile (shown in Fig.~\ref{fig:sb_shock}) are listed in the first two lines of Table~\ref{tab:shock_sb} and labelled ``Relic with notch''.
We first allowed all of the parameters to vary.
This fit is shown in the first line of Table~\ref{tab:shock_sb}, and in Figure~\ref{fig:sb_shock}.
The value of $\chi^2$ is somewhat high, and the deviations are due to both some inner points where it is likely that other
X-ray structures occur and the statistical errors are small, and to the outermost points where the statistics are not very good.
The emissivity exponents $p_i$ and $p_o$ correspond to the variation of the gas density with radius at a fixed temperature and abundances.
The two values are consistent within the errors.
When compared to a beta-model at large radii, the emissivity exponents are $ p = 3 \beta$.
Thus, the fitted exponents are consistent with $\beta \approx 1.1$, which is fairly large.
This indicates that the gas density is dropping rapidly with radius at this location, which is near the virial radius of the cluster.

The fitted value of the radius of the edge is very close to the outer radius of the radio relic, and the two appear to be consistent within the uncertainties.
To test this, we set $r_{\rm edge} = r_{\rm relic}$, fixed this parameter, and redid the fit.
This fit is shown in the second line of Table~\ref{tab:shock_sb}.
All of the parameters of this fit are consistent with those of the previous fit with a variable $r_{\rm edge}$ within the uncertainties.
Although the value of $\chi^2$/dof for this fit is slightly worse than that for the variable $r_{\rm edge}$, the
f-test indicates that the probability that the fit with a fixed $r_{\rm edge}$ is at least as good as the fit with a variable $r_{\rm edge}$ is 21\%, which is not terribly low.
Thus, the X-ray surface brightness edge is located at the outer edge of the radio relic to within the errors.

The best-fit values of $R$ are larger than unity by about 4-sigma.
As another test for the reality of the edge, we redid the fit, fixing the value of $r_{\rm edge} = r_{\rm relic}$, and fixing $R = 1$.
As shown in line 3 of Table~\ref{tab:shock_sb}, this is a very poor fit to the observed surface brightness profile.
The f-test indicates that the probability that there is no X-ray edge at the outer radius of the radio relic is only $2 \times 10^{-5}$.

As noted above, the NW radio relic has a ``notch'' at its southwest end which breaks the relatively smooth curve of the outer edge of the relic.
It is unclear whether this feature is due to a disturbance in the hydrodynamics of the ICM, or due to the geometry of pre-existing relativistic particles and magnetic field, or a projection effect (e.g., the notch is due to emission on the front or back side of a curved shock feature.
In an effort to eliminate these sources of systematic uncertainty, we also determined the radial X-ray surface brightness profile in a narrower elliptical wedge which excluded the notch (narrower solid lines in Fig.~\ref{fig:sb_regions}).
This wedge extended from an angle of $-$13.5$^\circ$ to $+$34$^\circ$ of the semi-major axis (measured counter-clockwise).
The results of a fit to the surface brightness profile in this wedge are shown in line 4 of Table~\ref{tab:shock_sb}.
This fit is significantly worse than the fit including the notch, and the uncertainties in the parameters are increased, in some cases very significantly. 
Except for the surface brightness, the edge parameters are consistent with the values of the whole wedge.
However, the X-ray surface brightness in the outer regions is lower, and the errors are thus much larger.
In particular, the uncertainties in the value of the $R$ are quite large;
this is the most important quantity in this fit, since it determines the compression and Mach number of the shock
(see \S\S~\ref{sec:discussion_shock} \& \ref{sec:shocksb} below).
However, the possibility that there is no edge (R $=$ 0) can be ruled out with high confidence (probability $< 1$\% by the f-test).

We also extracted the surface brightness profile for the notch region alone
(angles between $-$26$^\circ$ and  $-$13.5$^\circ$ of the semi-major axis;
Fig.~\ref{fig:sb_regions}).
This surface brightness profile showed the same shock feature at essentially the same radius
(the best fit radius was smaller by about 4\%).
However, because the area covered was about a factor of five smaller, the statistics were worse.
There also was a feature at the top of the Mushroom which is discussed in more detail below
(\S~\ref{sec:sb_mushroom}).
The outer edge of the Mushroom corresponds approximately to the inner edge
of the notch in the radio relic.
There was no clear feature at the outer edge of the radio relic, as might have been expected
if this edge also marked the position of a shock.

\begin{figure}[t]
 \begin{center}
 \includegraphics[scale=0.43, angle=0]{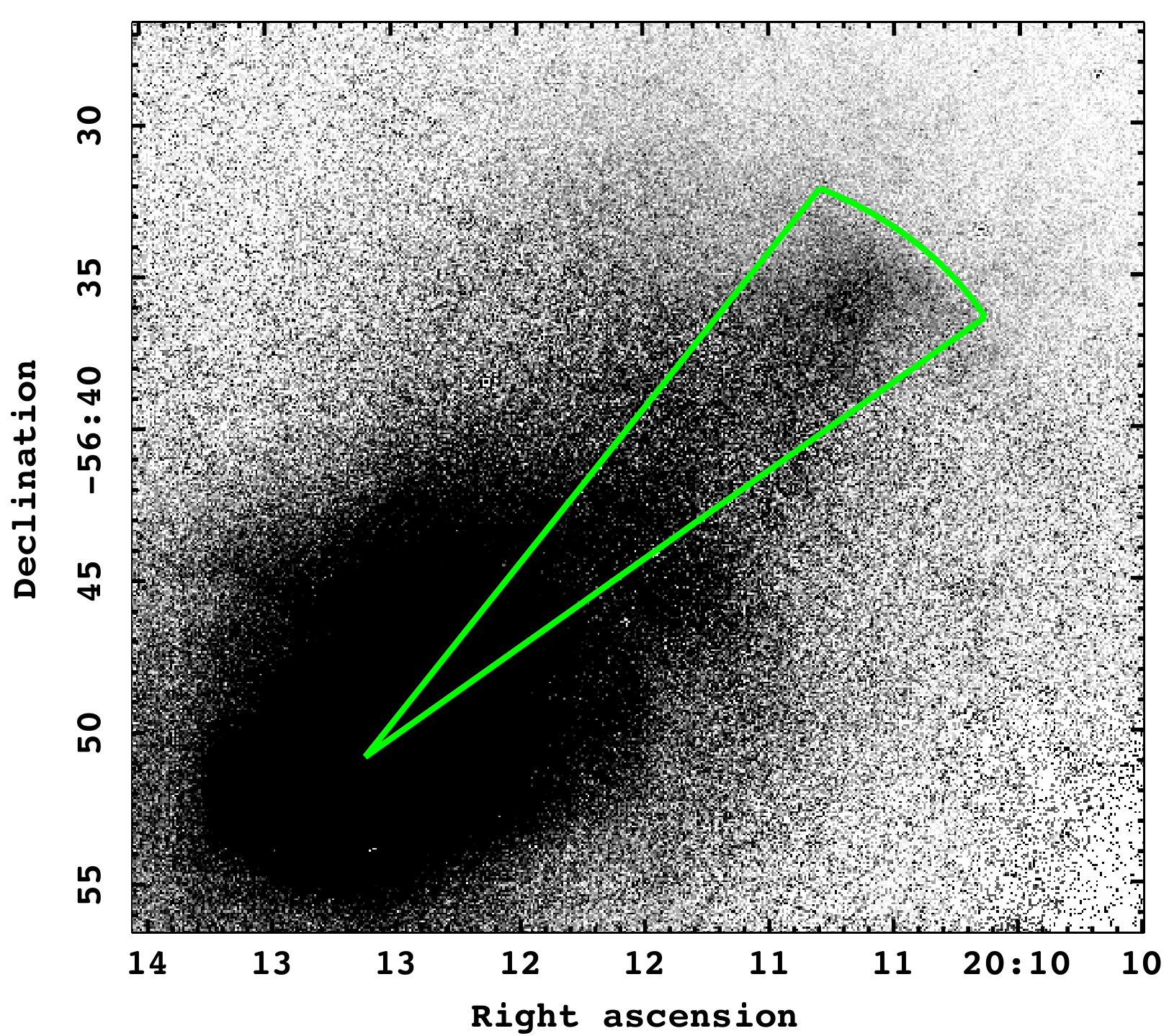}
 \end{center}
 \vskip-0.15in
 \caption{Shape of the elliptical annular wedges used to derive the radial X-ray surface brightness profile across the Mushroom feature.
 The greyscale image is the \xmmnewton\ X-ray image with the point sources removed.
 The solid green lines show the shape of the elliptical wedges used to derive the surface brightnesses.
 The radius of this elliptical wedge corresponds to the best-fit outer radius (semi-major axis of the ellipse) of
 1503\arcsec.
 The elliptical wedge starts at the cluster pressure peak (see Fig.~\ref{fig:sb_regions}).
 }
 \label{fig:sb_mushroom_regions}
\end{figure}

We also consider a set of broader elliptical pie annuli which include the extension of the radio relic to the NE (dashed lines in Fig.~\ref{fig:sb_regions}).
This was a considerably worse fit (line 6 in Table~\ref{tab:shock_sb}).
Moreover, the best-fit location of the edge was $\sim$36\arcsec\ inside of the outer edge of the radio relic.
This suggests that broadening the adapted outer limit of the radio relic to include the NE extension to the radio relic moved this radius beyond the actual X-ray edge at the two ends of the relic.
Also, the value of $R$ was poorly determined.
If the location of the edge was fixed to the the adopted outer radius of the radio relic, the fit was only slightly worse, but the value of $R$ was nearly unconstrained
(line 7 of Table~\ref{tab:shock_sb}).
The possibility that there was no edge could again be ruled out with high confidence (probability $< 0.3$\%), but the strength of the jump could be very large.
As noted above, $R$ is the most important parameter for the physical interpretation of the emission edge.
These results suggest that the broader region did not fit the shape of the X-ray edge as well as our standard ``Relic with notch'' region.
Thus, we will use the values in line 1 of Table~\ref{tab:shock_sb} as the standard parameters of the surface brightness edge.

Note that one key assumption in this analysis is that the principle axis of the emissivity edge is in the plane of the sky.
For other orientations, projection effects weaken the surface brightness discontinuity.
Thus, it is likely that the fitted values of $R$ underestimate the actual emissivity jump.

\begin{figure}[t]
 \begin{center}
 \includegraphics[scale=0.43, angle=0]{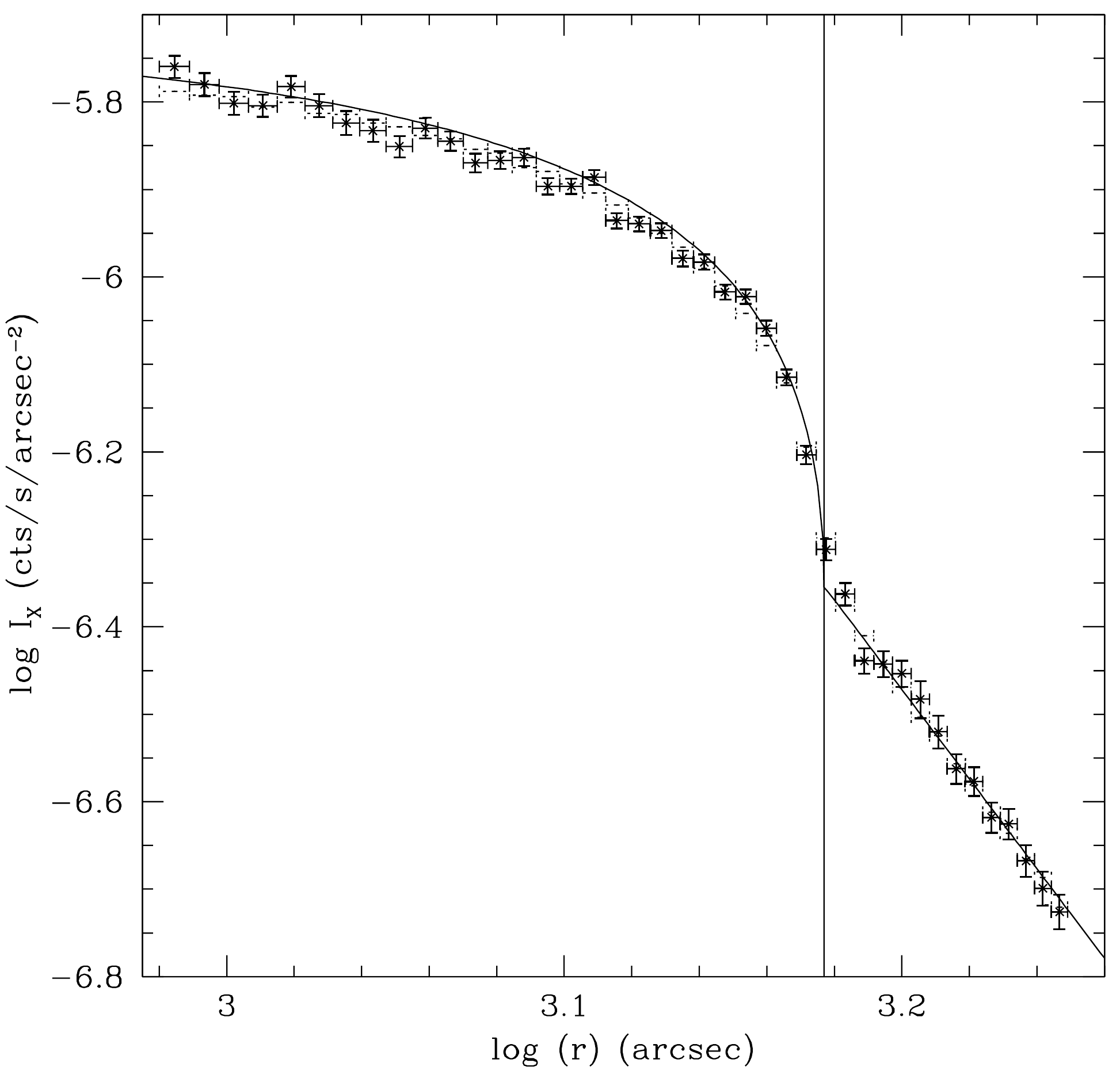}
 \end{center}
 \vskip-0.15in
 \caption{Radial X-ray surface brightness profile across the Mushroom feature in the 0.5--2.0 keV band, excluding the Al line at $\sim$1.5 keV.
 The profile was corrected for exposure and background, and point sources were removed.
 The notation is the same as in Fig.~\ref{fig:sb_shock}.
 The vertical line is the best-fit radius of the top of the Mushroom.
  An emissivity jump is evident at the top of the Mushroom.
 }
 \label{fig:sb_mushroom}
\end{figure}

\subsection{X-ray Surface Brightness Profile across the Mushroom}
\label{sec:sb_mushroom}

The radial surface brightness profile along the Mushroom feature was also extracted.
The shape of the elliptical wedge annular (epanda) regions used to accumulate the surface brightness in the 
region of the Mushroom is shown in Figure~\ref{fig:sb_mushroom_regions}.
The technique and model assumptions are the same as for the radial profile of the radio relic region
(\S~\ref{sec:sb_shock}),
The resulting surface brightness profile is given in Figure~\ref{fig:sb_mushroom}.

The surface brightness profile shows the characteristic shape associated with
a curved surface discontinuity in the X-ray emissivity.
The same model for an emissivity profile which has a discontinuity and is a power-law
function of radius was fit to this data.
The best-fit model is shown in Figure~\ref{fig:sb_mushroom} and
the parameters are given in row 8 of Table~\ref{tab:shock_sb}.
The best-fit emissivity jump is a factor of $2.75^{+0.25}_{-0.25}$.
This is a reasonably good fit.
There is clearly some more complicated structure in the surface brightness
profile, as is also shown in the image.
Note that the profile implies an inner X-ray emissivity which actually rises slightly
with increasing radius.
This may indicate that the region below the Mushroom is a tail of decreasing density, and
that the tail dominates the overall cluster radial surface brightness profile in this region.

\begin{figure}[t]
 \begin{center}
 \vskip-0.7in
 \includegraphics[scale=0.40, angle=0]{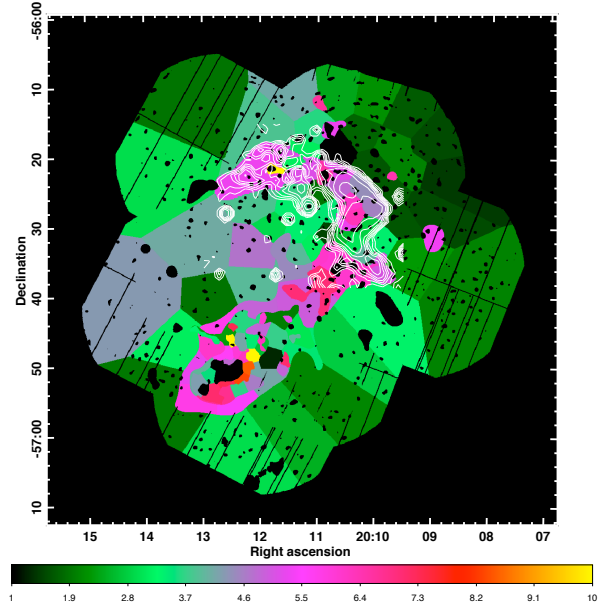}
 \end{center}
 \vskip-0.6in
 \caption{\xmmnewton\ temperature map of Abell 3667.
 The contours are from the SUMSS 843 MHz radio image the NW radio relic.
 The temperature map shows that the region of the radio relic is hot ($\sim5$ keV), while the region
 outside of the relic to the NW is much cooler ($\sim$2 keV).
 The temperature map also indicates that the Mushroom feature just to the south of the relic is also
 hot ($\sim$6 keV).
 }
 \label{fig:tmap}
\end{figure}

\section{X-ray Spectra}
\label{sec:spectra}

The spectra were fit in the energy range 0.8--12 keV, excluding the ranges from 1.35--1.75 keV and 7.5--9.9 keV which contain strong instrumental lines.
The foreground/background components included in the fit included the low energy Galactic foreground, particle background, and residual cosmic X-ray background from unresolved AGNs.
The upper end of the spectral band was included to provide better constraints on the residual particle and cosmic X-ray backgrounds.
The spectra were grouped to have at least 30 raw counts per bin to allow $\chi^2$ statistics to be applied.
Unless otherwise noted, we assume that the Galactic absorbing column was given by $N_H = 4.5 \times 10^{20}$ cm$^{-2}$,
and that the abundance of the cluster gas was 0.3 solar.

\subsection{Cluster Temperature Map}
\label{sec:spectra_tmap}

The resulting temperature map is shown in Figure~\ref{fig:tmap}.
All detected emission from sources with angular sizes of $\le$32\arcsec\ has been excluded from the fit;
these sources as well as the gaps in the pn coverage can be seen in Figure~\ref{fig:tmap} as black zones. 
In general, the outer parts of the cluster are relatively cool, although the errors are large in most directions due to the
relatively short exposures in the outer regions except for the NW radio relic region
(Fig.~\ref{fig:expmap}).
The region of the NW radio relic is relatively hot ($\sim5$ keV), while the region just outside the relic
to the NW is quite cool ($\sim$2 keV).
This is consistent with the presence of a shock at the outer edge of the radio relic.
Note also that the Mushroom region is fairly hot ($\sim$6 keV), which may indicate that this
feature is also due to a shock, rather than a cold front.

\begin{figure}[t]
 \begin{center}
 \vskip-0.1in
 \includegraphics[scale=0.43, angle=0]{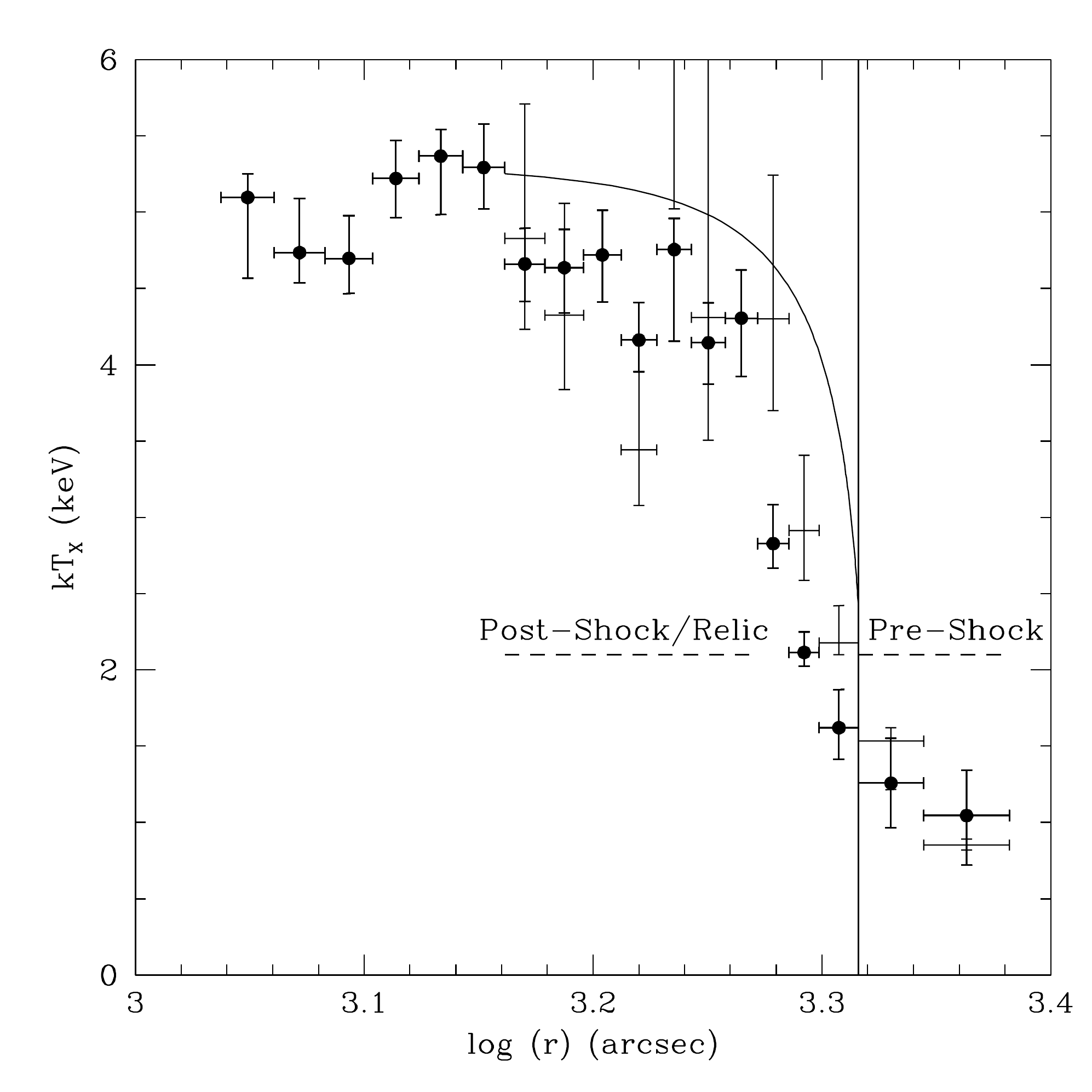}
 \end{center}
 \vskip-0.2in
 \caption{Filled circles give the best-fitted \xmmnewton\ temperatures in elliptical pie annuli regions which cross the NW radio relic as a function of the semi-major axis of the ellipse.
 The solid vertical line gives the approximate location of the outer edge of the NW radio relic.
 The widths of the error bars give the regions used to accumulate the spectra, while the vertical error bars are 1--$\sigma$ uncertainties.
 Note the rise in the temperature within the outer edge of the relic.
 The error bars without circles are de-projected temperatures for the outer 12 region fit using the {\sc projct} model in
 {\sc xspec}.
 The horizontal dashed lines show the radial regions used to accumulate the Post-Shock/Relic  and Pre-Shock spectra discussed in
 \S\S~\ref{sec:spectra_pre} and \ref{sec:spectra_post}.
 The thin curve shows the non-equipartition model for the post-shock temperature variation
 (\S\S~\ref{sec:discussion_tprofile} \& Appendix \ref{sec:equi}),
 which should be compared to the de-projected data points.
 }
\label{fig:tprofile}
\end{figure}

\subsection{Temperature Profile Across the Relic}
\label{sec:spectra_tprofile}

We determined the radial temperature profile across the region of the NW radio relic using elliptical pie annuli with the
same shape as those used to accumulate the surface brightness profile in \S~\ref{sec:sb_shock}.
Specifically, we used the standard panda regions in Figure~ \ref{fig:sb_regions} which included the `Notch' region.
The elliptical pie annuli were wider radially than those used to determine the SB profile in order to provide enough counts
for spectral fitting.
Starting at the ellipse which fit the outer edge of the radio relic, we chose epanda regions having a total number of at least 5000 source counts
in the 0.5--2.0 keV band.  
The large number was needed because of the relatively high ratio of background to source counts in these outer cluster regions.

Figure~\ref{fig:tprofile} shows the radial temperature profile (filled circle points) across the NW radio relic.
The temperature rises from $\sim$1.2 keV to $\sim$4.5 keV just inside of the outer edge of the relic.
This indicates that the X-ray emissivity jump seen at the same location (Fig.~\ref{fig:sb_shock}) is a shock.
Note, however, that the temperature doesn't rise discontinuously immediately at the outer edge of the relic,
but instead increases over a radial region of about 3 arcmin in width.
The best-fitted temperature is then approximately constant for $1450\arcsec \la r \la 1870\arcsec$.
At smaller radii, the temperature rises slightly again;
Figure~\ref{fig:tmap} indicates that this is mainly due to slightly hotter gas associated with the Mushroom feature.

In principle, the finite point-spread-function (PSF) of \xmmnewton\ will cause a sharp spectral change to appear more gradual.
However, the full-width-half-maximum for the \xmm\ mirrors and detectors is only around 6\arcsec, and the half-energy-width is only about 16\arcsec.
Thus, the PSF only makes a small contribution to the broadening to the temperature rise inside of the shock front.

\begin{deluxetable*}{lcccccccc}
\tablewidth{0pt}
\tabletypesize{\scriptsize}
\tablecaption{Pre-Shock and Post-Shock/Relic X-ray Spectral Fits
\label{tab:spectra}}
\tablehead{
&
\colhead{$r_{\rm in}$} &
\colhead{$r_{\rm out}$} &
&
\colhead{$k T$} &
\colhead{Norm$_T$\tablenotemark{b}}&
\colhead{$k T_2$ or $\Gamma$\tablenotemark{d}} &
&
\cr
\colhead{Region} &
\colhead{(arcsec)} &
\colhead{(arcsec)} &
\colhead{Model\tablenotemark{a}} &
\colhead{(keV)} &
\colhead{($10^{-3}$)} &
\colhead{(keV or none)} &
\colhead{Norm$_2$\tablenotemark{c}}&
\colhead{$\chi^2$/dof}
}
\startdata
Pre-Shock   & 2070 & 2410 & 1T    & $1.23^{+0.06}_{-0.17}$ & $0.214^{+0.007}_{-0.043}$ & &
& 2551.4/1674 $=$ 1.524 \\[4pt]
Post-Shock & 1450 & 1870 & 1T    & $4.37^{+0.93}_{-0.76}$ & $1.403^{+0.039}_{-0.038}$ & &
& 2237.9/1639 $=$ 1.365 \\[4pt]
                  &          &          & 2T     & $5.35^{+2.23}_{-1.26}$ & $1.263^{+0.026}_{-0.026}$
& (1.23) & $0.109^{+0.029}_{-0.025}$ & 2200.1/1638 $=$ 1.343  \\[4pt]
                  &          &          & 2T     & $5.68^{+1.69}_{-1.13}$ & $1.226^{+0.052}_{-0.036}$
& (1.23) & (0.1335) & 2202.6/1639 $=$ 1.344  \\[4pt]
                  &          &          & 2T     & $5.58^{+2.40}_{-1.39}$ & $1.244^{+0.024}_{-0.021}$
& $0.95^{+0.02}_{-0.02}$ & $0.258^{+0.036}_{-0.045}$ & 4730.7/3312 $=$ 1.428  \\[4pt]
                   &          &         & 1TPL & $6.11^{+3.51}_{-1.65}$ & $1.213^{+0.035}_{-0.064}$
& $5.21^{+0.45}_{-0.74}$  & $1.35^{+0.43}_{-0.19}$ & 2195.5/1637 $=$ 1.341  \\[4pt]
                   &          &         & 1TPL & $8.33^{+7.43}_{-4.50}$ & $0.497^{+0.188}_{-0.079}$ & (2.10)                             & $2.45^{+0.38}_{-0.59}$ & 2210.1/1638 $=$ 1.349  \\[4pt]
\enddata
\tablenotetext{a}{1T $\equiv$ single temperature;
2T $\equiv$ two temperatures; 
1TPL $\equiv$ single temperature plus power-law}
\tablenotetext{b}{Normalization of the APEC thermal spectrum,
which is given by $\{ 10^{-17} / [ 4 \pi (1+z)^2 d_A^2 ] \} \, \int n_e n_H
\, dV$, where $z$ is the redshift, $d_A$ is the angular diameter distance,
$n_e$ is the electron density, $n_H$ is the ionized hydrogen density,
and $V$ is the volume of the region.
Note that this is 0.001 times smaller than the conventional normalization in {\sc xspec}.}
\tablenotetext{c}{The normalization of the second spectral component.
If this is a thermal component, the units are the same as for the first component\tablenotemark{b}.
If the second component is a power-law, the normalization is the
photon flux at 1 keV in units of
$10^{-4}$ photons cm$^{-2}$ s$^{-1}$ keV$^{-1}$.}
\tablenotetext{d}{Values in parenthesis are fixed during the fit.}
\end{deluxetable*}

The accumulated spectra include emission along the entire line-of-sight, and thus contain emission from gas at
larger radii.
We have determined the best-fitted temperatures by de-projecting these spectra.
In doing this, we made the same assumptions used to de-project the X-ray SB profile to give the emissivity in
\S~\ref{sec:sb_shock}:
that the temperature is constant on aligned self-similar prolate spheroidal surfaces whose shape is given by the outer
edge of the NW radio relic, and that the major axis of the prolate spheroid is in the plane of the sky.
The error bars without filled circles show the best-fitted de-projected temperatures determined using the
{\sc projct} function in {\sc xspec}.
We also did the de-projection using the ``onion-peeling'' algorithm discussed in \citet{BSM03}, which is more robust and
stable when there are large gradients in the X-ray SB.
Both techniques gave very similar results.
While projection can account for a non-trivial portion of the broadening, the temperature still
increases over a radius of about 2\arcmin.
The effect of projection will increase if the major axis of the shock front is not in the plane of the sky, and this might
account for the broadening.
More generally, other breakdowns in the simple ellipsoidal X-ray emissivity model we have assumed might have the
effect of broadening the observed temperature profile.
However, this would also affect the SB profile, which is reasonably well-fitted by the simple projection model 
(Table~\ref{tab:shock_sb} and Fig.~\ref{fig:sb_shock}).
The absence of large gradients in the radial velocity distribution of the galaxies also suggests that
the merger is nearly in the plane of the sky
\citep{Joh-Hol04}.

The broadening of the temperature profile at the outer edge of the relic may also be due to inhomogeneities in the temperature structure of the gas, either on the plane of the sky (e.g., the `Notch' feature in the relic) or along the line-of-sight.
Possible physical explanations for the temperature profile are discussed in \S~\ref{sec:discussion_tprofile}.

\subsection{Pre-Shock X-ray Spectrum}
\label{sec:spectra_pre}

We also determined the X-ray spectrum of the gas just beyond and just within the outer edge of the NW radio relic.
For this purpose, we used larger regions to provide a more accurate determination of the temperature and other properties.
The resulting fits are summarized in Table~\ref{tab:spectra}.

For the gas just beyond the relic (the ``Pre-Shock'' gas), we fit the  spectrum in an elliptical pie annulus extending from 2070\arcsec\ to
2410\arcsec\  (Fig.~\ref{fig:tprofile}).
While this gave a fit that agreed well with the values in Figure~\ref{fig:tprofile}, the $\chi^2$ value was rather high,
suggesting that there are remaining systematic uncertainties in the background.
The regions used to accumulate the pre- and post-shock spectra are large enough that the statistical errors
in the spectra are fairly low.
But, given the low cluster surface brightness at these large radii, systematic errors in the background are
likely to be important.
Therefore, we included a systematic error of 3\% in the background correction when determining the errors
in the fit parameters.
The values of $\chi^2$ include only the statistical errors.
This gave a pre-shock temperature of $kT = 1.23^{+0.06}_{-0.17}$ keV.
If the abundance was allowed to vary, the best fit value was $0.075 \pm 0.013$, but the fit was not improved significantly.
This is a rather low abundance, but perhaps not unreasonable for gas $\sim$2.3 Mpc from the cluster center.

\subsection{Post-Shock/Relic X-ray Spectrum}
\label{sec:spectra_post}

The spectrum of the post-shock gas was extracted from an elliptical pie annulus extending from 1450\arcsec\ to
1870\arcsec\  (Fig.~\ref{fig:tprofile}).
The outer edge of this region is well within the apparent shock location from the X-ray surface brightness profile or the outer edge of the radio relic.
The reason for this choice is the rather extended rise in the temperature as seen in the temperature profile.
This rise might be due in part to the finite spatial resolution of \xmmnewton,
although the angular region of the rise is too broad for this to be the main cause.
It is partially due to projection effects, although Figure~\ref{fig:tprofile} suggests that this is not the
sole cause.
It may be due to temperature inhomogeneities due to structure in the shock front, such as the Notch.
It may have another astrophysical cause (\S~\ref{sec:discussion_tprofile}), such as electron-ion non-equipartition at
the shock.
In any case, it seemed safer to exclude the transition region, and extract the post-shock spectrum from the region after the temperature
profile has flattened.
This post-shock region still includes 59\% of the radio flux of the relic.

\subsubsection{Post-Shock Thermal Emission}
\label{sec:spectra_post_thermal}

\begin{figure}
 \begin{center}
 \includegraphics[scale=0.35, angle=-90]{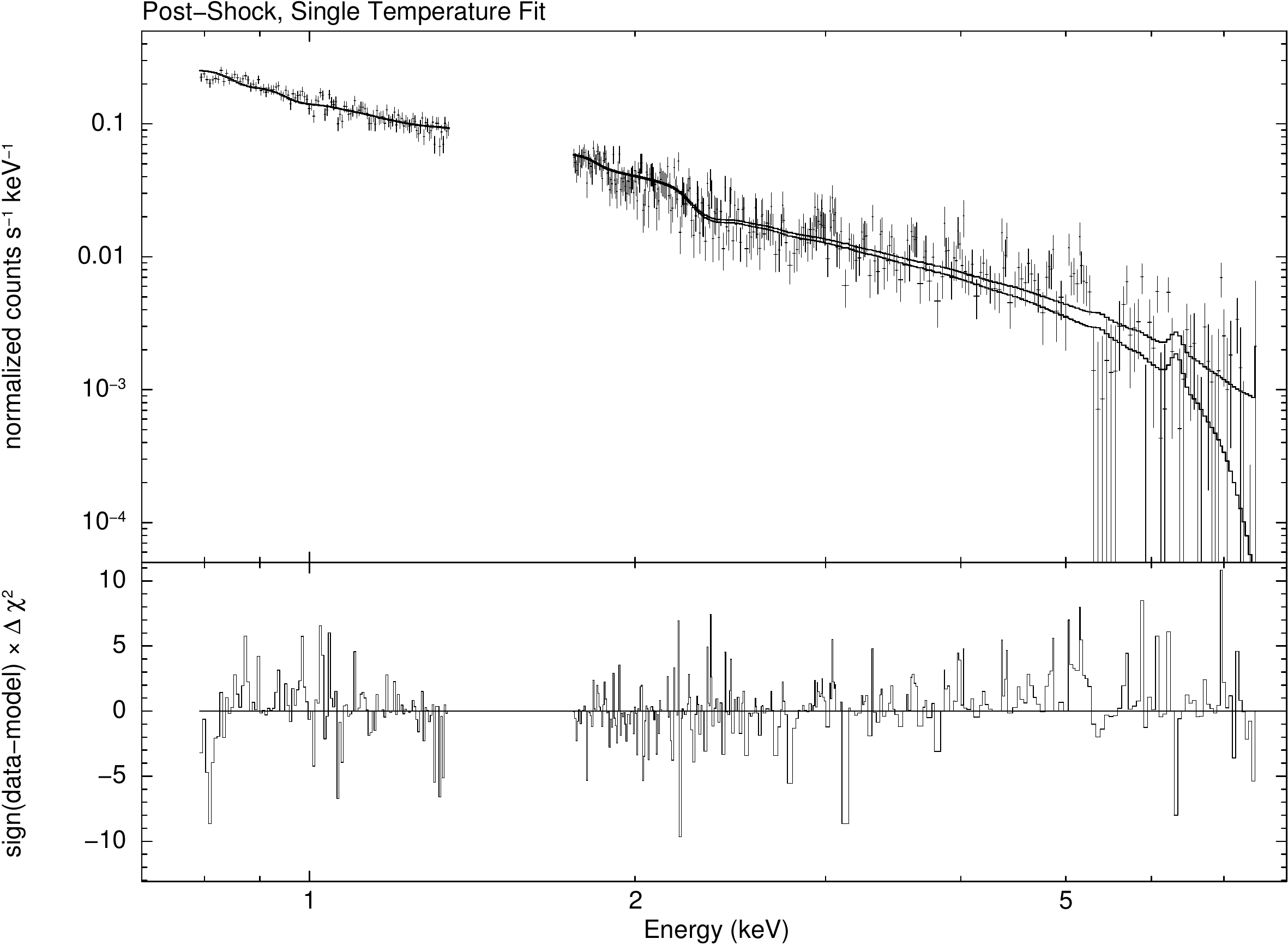}
 \end{center}
 \vskip-0.15in
 \caption{Best-fit single temperature model for the post-shock spectrum (row 2 in Table~\ref{tab:spectra}).
 The upper panel gives the data and the model.
 The lower solid curve is the model for the cluster emission, while the upper curve gives the total model including background.
 The lower panel gives the residuals to the fit, in terms of the contribution to $\chi^2$, multiplied by the sign of the residual.
 }
 \label{fig:tpostshock}
\end{figure}

The post-shock spectrum was fit initially with a single temperature model.
The best-fit model had a temperature of $kT = 4.37^{+0.93}_{-0.76}$ keV
(row 2 in Table~\ref{tab:spectra}), which is consistent with the results from the temperature profile
(Fig.~\ref{fig:tprofile}).
This fit is shown in Figure~\ref{fig:tpostshock}.
If the abundance is allowed to vary, the best fit value is $0.08 \pm 0.03$ solar,
which is low but consistent with the pre-shock spectral fits.
If the abundance is fixed at 0.3 solar, the temperature is $kT = 4.80^{+0.20}_{-0.13}$ keV,
and the fit is only slightly worse.

We next tried a two-temperature fit to the spectrum.
However, the temperatures were not well constrained.
One motivation for a two temperature model is that some of the pre-shock gas is almost certainly projected onto the post-shock region.
Thus, we also tried a two temperature fit
in which the outer thermal component was fixed to the spectrum of the pre-shock gas
(row 1 in Table~\ref{tab:spectra}).
Initially, the normalization of this second, cooler component was allowed to vary.
This resulted in a higher temperature ($kT = 5.35^{+2.23}_{-1.26}$ keV;
row 3 in Table~\ref{tab:spectra}) as expected.
The flux of the best-fit cool component was reasonably close to what one might expect from projection,
if one assumes that the geometry and gas density distribution were given by the best-fit X-ray surface brightness model near the shock (\S~\ref{sec:sb_shock} and Table~\ref{tab:shock_sb}), and one assumes that all of the pre-shock gas has the same spectrum.
The surface brightness model predicts that the normalization of the cool gas projected on the post-shock spectral regions is 0.62 times the normalization in the pre-shock spectral region.
This gives a normalization of 0.1335 (in the units in Table~\ref{tab:spectra}), whereas the best-fit value is
0.109.

We also fixed the spectrum and normalization of the cool component in the post-shock regions to the spectrum of the pre-shock gas and the normalization predicted by the X-ray surface brightness model
(row 4 in Table~\ref{tab:spectra}).
This was only a slightly worse fit with a consistent temperature $kT = 5.68^{+1.69}_{-1.13}$ keV.
This result corresponds to the ``onion-peeling'' spectral de-projection algorithm discussed in \citet{BSM03}.

Finally, we fit the pre- and post-shock spectra simultaneously, including the pre-shock gas in the projected
post-shock spectra based on the model for the X-ray surface brightness
(row 5 in Table~\ref{tab:spectra}).
This gave a consistent value for the de-projected post-shock temperature
($kT = 5.58^{+2.40}_{-1.39}$ keV), but a considerably lower temperature for the pre-shock gas.
Because we believe it is safer to determine the temperature of the pre-shock gas purely from a region containing only that gas, we prefer the previous fit
(row 1 in Table~\ref{tab:spectra}).
However, the {\sc xspec} fit gave a value for the covariance between the pre- and de-projected post-shock temperatures of
$\sigma(T_1,T_2) = -6.00 \times 10^{-6}$ keV$^2$, which is useful for the subsequent analysis.

\subsubsection{Post-Shock IC Emission}
\label{sec:spectra_post_ic}

The same relativistic electrons which produce the synchrotron radio emission of the radio relic will also produce X-ray emission by inverse Compton (IC) scattering of Cosmic Microwave Background photons.
For reasonable values of the magnetic field, X-ray emission in the \xmmnewton\ band will be produced by lower energy relativistic electrons than those which generate the radio emission in the observed images.
Thus, the post-shock region may also have some IC emission, as well as the thermal emission from the shocked intracluster gas.
To search for this emission in the post-shock spectrum, we have included a power-law spectral component in our models.
Initially, we allowed both the photon spectral index $\Gamma$ and the normalization of the power-law component to vary during the fit.
This results in a power-law spectrum with an unphysical steep spectral index,
$\Gamma = 5.21^{+0.45}_{-0.74}$
(row 7 in Table~\ref{tab:spectra}).
This power-law component only affects the softest X-ray channels in the spectrum.
The fitted spectral index is much higher than that of the radio relic synchrotron spectrum.
For synchrotron and IC emission from the same relativistic electrons, the radio and X-ray spectral indices should agree.
Since the IC X-ray emission here should actually come from somewhat lower energy electrons than the radio emission, one might expect the
IC spectral index to actually be a bit smaller than the radio value.
Thus, we fixed the IC spectral index the observed radio value of $\Gamma = 2.1$ \citep{RWH+97}.
This led to a fit
(row 8 in Table~\ref{tab:spectra})
in which roughly half of the flux comes from the non thermal component.
However, the temperature of the thermal ICM is then very high ($\sim$8 keV, but with very large uncertainties), which might be inconsistent with
the intracluster gas temperatures at slightly smaller radii or with the jump conditions for a shock with a
Mach number consistent with the gas density jump.
The 1TPL fit is better than the 1T fit, but only slightly.

Another issue with a large IC contribution to the post-shock spectrum is the slow rise in temperature behind the shock
(Fig.~\ref{fig:tprofile} and \S\S~\ref{sec:spectra_tprofile} \& \ref{sec:discussion_tprofile}).
For a power-law energy distribution of relativistic electrons, the radio synchrotron and X-ray IC emission have the same power-law spectral shape.
The radio spectrum immediately behind the shock is fairly flat \citep{Joh-Hol04},
which suggests the IC spectrum should also be quite hard.
The relic radio spectrum steepens with distance from the shock, and this is consistent with energy
losses by the relativistic electrons behind the shock
(\S~\ref{sec:discussion_loss}).
Thus, if anything, the X-ray IC spectrum should become softer further behind the shock.
But, the observed spectra actually gets harder (Fig.~\ref{fig:tprofile}).
This suggests that the majority of the post-shock X-ray emission is not due to IC.

Given these issues, we treat the flux of the possible IC component at an upper limit.
This gives an upper limit on the IC photon flux of $< 3.2 \times 10^{-4}$ cts cm$^{-2}$ s$^{-1}$ or 
an energy flux of $< 1.0 \times 10^{-12}$ erg cm$^{-2}$ s$^{-1}$ (0.8--8.0 keV, unabsorbed, 90\% confidence level).
Since the post-shock spectral region includes only 59\% of the radio relic flux, the limit of the total IC flux
would be 
$< 5.4 \times 10^{-4}$ cts cm$^{-2}$ s$^{-1}$
($< 1.7 \times 10^{-12}$ erg cm$^{-2}$ s$^{-1}$)
assuming the IC X-ray flux is proportional to the radio flux.
This limit is similar to previous limits \citep[][\citetalias{FSN+10}]{Nak+09}, and implies a lower limit on the magnetic field in the relic of
$B \ga 3$ $\mu$G.
The upper limits from the lack of Faraday rotation are similar \citep{Joh-Hol04}, which suggests that $B \sim 3$ $\mu$G.
We will assume this value in further discussions.

\subsection{Mushroom Spectra}
\label{sec:spectra_musroom}

\begin{figure}[t]
 \begin{center}
 \includegraphics[scale=0.43, angle=0]{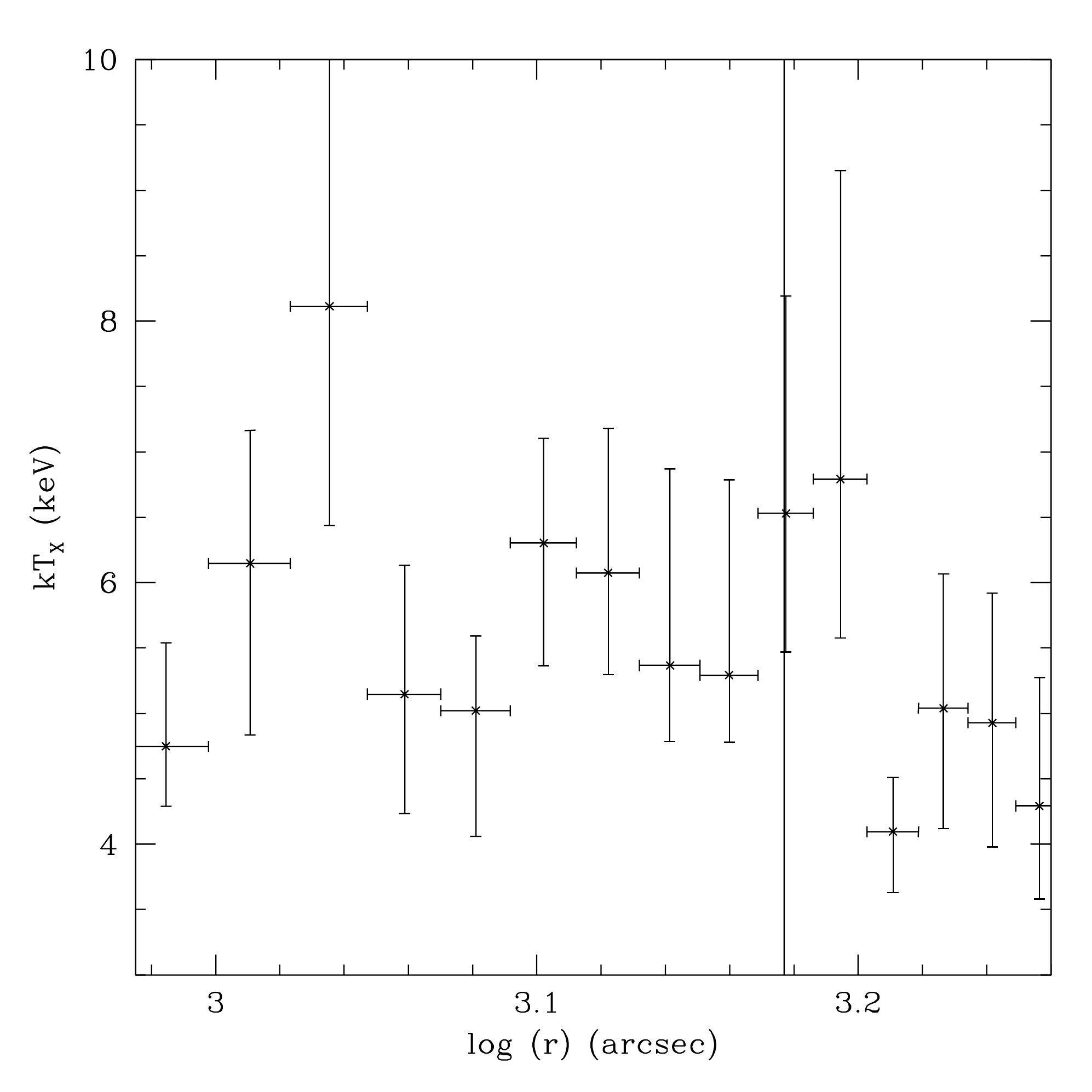}
 \end{center}
 \vskip-0.2in
 \caption{The radial temperature profile across the Mushroom feature.
 The solid vertical line gives the best-fit radius of the top of the Mushroom as determined
 from the radial surface brightness profile.
 The widths of the error bars give the regions used to accumulate the spectra, while the vertical error bars are 1--$\sigma$ uncertainties.
 No very clear temperature trend is obvious.
  }
\label{fig:mushroom_tprofile}
\end{figure}

The radial temperature profile of the gas in the Mushroom was determined from spectra accumulated in epanda regions with the same
shape as those used to determine the radial X-ray surface brightness profile
(\S~\ref{sec:sb_mushroom}).
Wider annuli were used for the spectra since more counts are needed to derive temperatures.
The techniques were the same as those used to derive the radial temperature profile across
the radio relic region
(\S~\ref{sec:spectra_tprofile}).
The resulting temperature profile is shown in Figure~\ref{fig:mushroom_tprofile}.
While there is a suggestion that the gas just above the Mushroom is hotter than
the gas within it, the trend is not very clear.

\section{Discussion}
\label{sec:discussion}

\subsection{Shock Jump Conditions}
\label{sec:discussion_shock}

Both the surface brightness profile (Fig.~\ref{fig:sb_shock}) and the temperature profile
(Fig.~\ref{fig:tprofile}) show a discontinuity at the location of the outer edge of the radio
relic, which indicates that a shock is present there.
The jump in the X-ray emissivity and the jump in temperature each provide a nearly independent
estimate of the shock Mach number ${\cal{M}}$.

The jump in the X-ray emissivity at the shock found from the X-ray surface brightness profile is
$R = 3.11^{+1.75}_{-0.94}$
(\S~\ref{sec:sb_shock}, Table~\ref{tab:shock_sb}).
This ratio must be corrected for the temperature-dependent emission function of the gas
(Eq.~\ref{eq:compress}).
Following the procedure in Appendix~\ref{sec:shocksb}, we find
$( \Lambda_2 / \Lambda_1 ) = 0.57^{+0.04}_{-0.10}$.
This gives a shock compression $ C \equiv \rho_2 / \rho_1 = 2.34^{+0.69}_{-0.36}$,
where $\rho$ is the gas mass density, and the subscripts 1 and 2 refer to the pre- and post-shock
gas, respectively.
The shock Mach number is then ${\cal{M}} = 2.05^{+0.73}_{-0.38}$.

The pre- and post-shock temperatures are
$k T_1 = 1.23^{+0.06}_{-0.17}$ keV
(row 1 in Table~\ref{tab:spectra})
and
$k T_2 = 5.35^{+2.23}_{-1.26}$ keV (row 3 in Table~\ref{tab:spectra}), respectively.
For the post-shock temperature, we use the value de-projected by fitting the projected spectrum
to the pre-shock spectral shape (with a variable normalization) plus the post-shock spectrum.
As noted in \S~\ref{sec:spectra_post_thermal}, the normalization of the pre-shock spectrum
in this fit is within 20\% of the value predicted by the best-fit X-ray surface brightness profile
(\S~\ref{sec:sb_shock}).
But, fitting the normalization from the observed post-shock spectra should help to insure against
small differences in the pre-shock gas distribution along the line-of-sight to the post-shock spectral
region.
The effects of projection are somewhat reduced by the fact that our chosen post-shock spectral region is displaced inwards from the shock front, and the X-ray surface brightness profile declines rapidly.
This gives a temperature ratio of
$( T_2 / T_1 ) = 4.35^{+1.91}_{-1.05}$.
The shock jump conditions then imply a Mach number of
${\cal{M}} = 3.34^{+0.91}_{-0.50}$.
The Mach number from the temperature jump is 1.46 $\sigma$ higher than the value from the X-ray surface brightness jump.
The mean shock number from these two values weighted by their errors is
${\cal{M}} = 2.54^{+0.80}_{-0.43}$, which we adopt as our best value for further analysis.
Although the uncertainties are large, this Mach number makes the NW Relic shock in Abell 3667
one of the strongest shocks seen in a cluster,
although weaker than that in the Bullet Cluster
\citep[${\cal{M}} = 3.0 \pm 0.4$;][]{MV07}.

This Mach number implies a shock compression of
$C = 2.73^{+0.55}_{-0.29}$,
and a temperature jump of
$( T_2 / T_1 ) = 2.86^{+1.29}_{-0.69}$.
The pre-shock sound speed is
$c_s = 569^{+14}_{-39}$ km s$^{-1}$, which implies a
shock speed of
$v_s = {\cal{M}} c_s = 1450^{+460}_{-260}$ km s$^{-1}$.
The speed of the post-shock gas relative to the shock is
then
$v_2 = v_s / C = 530^{+60}_{-50}$ km s$^{-1}$.

Our fit to the X-ray surface brightness also allows the pre- and post-shock densities to be determined.
Our best-fit model gave a surface brightness at the shock apex of
$ I_o = (2.61^{+0.67}_{-0.53}) \times 10^{-8}$ cts s$^{-1}$ arcsec$^{-2}$
(Tab.~\ref{tab:shock_sb}).
This implies a pre-shock electron density of
$n_{e1} = 5.51 \times 10^{-5}$ cm$^{-3}$ from
equations~(\ref{eq:Io}) and (A12) in \citet{Kor+11}.
The post-shock electron density is $n_{e2} = C n_{e1} = 1.51 \times 10^{-4}$ cm$^{-3}$.

The measured values of the density and temperature jump at the shock are somewhat inconsistent
for a $\gamma = 5/3$ shock.
The temperature jump is larger than expected given the density jump, which could mean that the gas is
``less isothermal'', which suggests that the effective adiabatic index is $\gamma_{\rm eff} > 5/3$.
If one assumes that the shock compression and temperature jump are consistent
(\S~\ref{sec:gamma}),
the implied effective adiabatic index and shock Mach number are
$\gamma_{\rm eff} = 2.05^{+0.32}_{-0.52}$ and ${\cal{M}} = 2.80^{+0.76}_{-0.50}$.
If the shocked gas contained a significant energy density of relativistic particles or turbulent magnetic fields,
one might expect a value of $\gamma_{\rm eff} \approx 4/3$, which is the opposite trend from what is observed.
However, if there is a strong, ordered magnetic field, particularly parallel to the shock front,
then one would expect
$\gamma_{\rm eff} \approx 2$, which is close to the required value
\citep{Hel53}.
Thus, the shock jump conditions might provide some evidence for a dynamically important, ordered magnetic field with a significant component parallel to the shock front.
This would also be consistent with the the observed radio polarization of the radio relic just behind the
shock front
\citep{Joh-Hol04}.

The measured X-ray surface brightness and temperature jump were fit simultaneously to a shock model with a magnetic field which is parallel to the shock front (\S~\ref{sec:magnetic}).
This model had a ratio of magnetic to gas pressure of $b \approx 3.55$,
and a shock Mach number of ${\cal{M}} \approx 5.12$,
giving a shock velocity of $v_s \approx 2920$ km s$^{-1}$.
Although not ruled out by present observations, the magnetic field strength and shock speed are both rather large.
Thus, with the fact that the temperature jump is larger than expected for the density jump in Abell 3667 would be consistent with a dynamically non-trivial magnetic field aligned roughly parallel to the shock front, the large errors make any conclusion uncertain.
Additional deep observations of other merger shocks would be useful to see if there is a consistent pattern of large temperature jumps relative to the X-ray surface brightness jumps.

Since the difference in Mach numbers from the density and temperature jump are only mildly inconsistent, and the parameters of the ordered magnetic field model seem too extreme, we will continue to use the average shock properties from the temperature and density jump in the subsequent analysis.

\subsection{Temperature Profile at Shock}
\label{sec:discussion_tprofile}

As noted in \S~\ref{sec:spectra_tprofile} and Figure~\ref{fig:tprofile},
the temperature increase within the shock is more gradual than might have been
expected given the \xmmnewton\ PSF.
Much of the slow rise in temperature can be explained by the projection of pre-shock gas
in front of and behind the shock (the lighter error bars without circles in Fig.~\ref{fig:tprofile}).
However, the de-projected temperature profile still requires about 2\farcm3 to rise from the pre-shock value to a plateau at the post-shock value.

One simple explanation for the slow rise in the temperature would be inhomogeneities in the gas near the shock surface, either along our line-of-sight or in the plane of the sky.
One obvious inhomogeneity is the Notch at the western end of the radio relic (e.g., Fig.~ \ref{fig:sb_regions}).
This feature extends roughly 5\arcmin\ within the assumed elliptical shock surface.
The temperature map (Fig.~ \ref{fig:tmap}) does suggest that the gas in this region is cooler than the remainder of the shock.

Alternatively, it is uncertain how effective collisionless astrophysical shocks are at heating electrons.
If the electrons are not heated very effectively in the shock, then
given the very low gas densities in this region roughly 2.2 Mpc from the
center of the cluster and the high shock speed, it is possible that the
post-shock gas has not had time to come into electron-ion equipartition by
Coulomb collisions, or into collisional ionization equilibrium.
If the electrons are not strongly heated by the shock, the time scale for
the electron and ion temperatures to come into equipartition is
approximately
\citep{FL97,WS09}
\begin{equation}
t_{\rm eq} \approx 6.3 \times 10^7
\left( \frac{T_e}{10^7 \, {\rm K}} \right)^{3/2}
\left( \frac{n_p}{10^{-4} \, {\rm cm}^{-3} } \right)^{-1}
\left( \frac{\ln \Lambda}{40} \right)^{-1}
\, {\rm yr}
\, ,
\label{eq:teq}
\end{equation}
where $T_e$ is the electron temperature, $n_p$ is the proton density, and
$\ln \Lambda$ is the Coulomb logarithm. 
Because the merger shock
has a relatively low Mach number (as is true of all merger shocks),
the electrons will be heated significantly
by adiabatic compression, even if there were no shock electron heating.
For the adapted value for the shock compression, adiabatic heating gives
$(T_{e2} / T_{e1} ) \approx 1.95^{+0.26}_{-0.14}$,
while the full shock heating including adiabatic compression gives
$( T_2 / T_1 ) = 2.86^{+1.29}_{-0.69}$.
If the adapted
post-shock density and the post-shock electron temperature assuming only
adiabatic heating are used in equation~(\ref{eq:teq}), the approximate time
to reach equipartition is $t_{\rm eq} \approx 2.4 \times 10^8$ yr.
Thus, the thickness of the region with a lowered electron
temperature would be $d_{\rm eq} \approx v_2 t_{\rm eq} \approx 0.13$ Mpc,
corresponding to an angular scale of $\theta_{\rm eq} \approx 2.1 \cos \phi$
arcmin, where $\phi$ is the angle between the central shock normal and the plane of the sky.
Combined with the effects of projection (Fig.~\ref{fig:tprofile}), non-equipartition could
explain the width of the region over which the observed X-ray spectral temperature rises
within the shock.

In Figure~\ref{fig:tprofile}, the narrow curve gives the expected variation in the electron temperature behind
the shock due to non-equipartition.
The assumptions are
that the total electron plus ion temperature (the total thermal energy) behind the shock
is constant and given by the de-projected temperature from the post-shock spectrum
(row 3 of Table~\ref{tab:spectra}),
that the gas density increases as an inverse power-law of radius (Tab.~\ref{tab:shock_sb}),
and that our best-fit values of $n_{e2}$, $T_2$, $T_1$, shock compression $C$, and shock
speed $v_s$ are all correct, and that the apex of the shock lies in the plane of the sky.
Given that there are no free parameters in this Coulomb heating model, it may not be surprising that
this is not a very good fit.
However, its biggest flaws are the jump in electron temperature at the shock due to adiabatic
compression, and the subsequent rapid increase in the temperature within the shock.
The overall rapid increase in the temperature in the model compared to the de-projected spectral
fits suggest that non-equipartition is not the primary cause of the slow temperature rise.

Another result of the low gas density and high shock speed in the relic shock is a departure
from collisional ionization equilibrium.
Unlike non-equipartition (which will not be important if electrons are effectively heated in the shock) or
transport processes such as thermal conduction (which can be greatly reduced in importance by magnetic effects), non-equilibrium collisional ionization is certain to occur in low density shocks.
However, if the pre-shock gas temperature is high, most of the X-ray emission is due to thermal bremsstrahlung rather than line emission, and the spectral signatures of non-equilibrium ionization may be subtle.
In the Abell~3667 relic shock, the initial gas temperature is fairly low
($kT_1 \approx 1.23$ keV, Table~\ref{tab:spectra}), and X-ray line emission should be more important,
and non-equilibrium ionization may affect the overall spectrum.
At the observed post-shock temperature in Abell~3667, the time required to achieve
equilibrium ionization is about $t_{\rm ioneq} \sim 3 \times 10^{12} /
n_e$ sec \citep[e.g.,][]{Fuj+08}.
At the post-shock density, this gives $t_{\rm ioneq} \sim 6 \times 10^8$ yr, which is
very similar to the time to establish equipartition.

We checked to see if non-equilibrium ionization affected the temperature variation
just within the shock in Figure~\ref{fig:tprofile}.
First, we simulated non-equilibrium ionization X-ray spectra for the observed regions in Figure~\ref{fig:tprofile}
assuming that the pre-shock gas was in ionization equilibrium at $k T_1 = 1.23$ keV, that the electron temperature in the post-shock gas as $k T_1 = 5.35$ keV, that the abundance was 0.3 Solar, and that the ionization time-scale parameter $\tau \equiv \int n_e d t$ for each region was determined by integrating the post-shock density model from our X-ray surface brightness fits and the flow timescale given by $v_2$ to the center point of each region.
The spectra were simulated in {\sc xspec}%
\footnotemark[7]
using the RNEI model and the \xmmnewton\ instrument responses for that region.
Then, these simulated non-equilibrium  spectra were fit with the APEC model in exactly the same way as the real spectra.
We found that the temperature values were only reduced very slightly below $kT_2 = 5.35$ keV, and were still well within the errors in the spectral fits.
This is due to the fact that the temperature was mainly affected by the shape of the X-ray continuum at an electron temperature of 5.35 keV.
The main effect of non-equilibrium ionization was to strengthen the Fe K lines.
Thus, the single temperature equilibrium fit to the non-equilibrium simulated spectra gave higher iron abundances, while the real spectral fits gave low abundances.
(If IC emission contributes to the post-shock spectrum, this would reduce the apparent abundance of iron
behind the shock.)

Our second test was to de-project the observed spectra in Figure~\ref{fig:tprofile} and fit them with a non-equilibrium RNEI model, but one in which only the overall normalization and the ionization scale parameter $\tau$ were allowed to vary.
These fits all gave large values of $\tau$, implying that the gas was in ionization equilibrium.
Moreover, the $\tau$ values did not increase with distance from the shock as expected.

Thus, the conclusion is that the slow increase in the post-shock temperatures in Figure~\ref{fig:tprofile}
is probably not due to non-equilibrium ionization.
It remains of bit of a mystery why non-equilibrium ionization does not seem to have increased the
strength of the Fe K lines, though.
However, If the iron abundance is as low as allowed by the pre-shock spectrum, the non-equilibrium model is still marginally consistent with the observed spectrum.

Another possibility is that the shock energy is initially dissipated into some mix of thermal and nonthermal energy, and the nonthermal energy decays into thermal energy in the post-shock region.
For example, part of the shock energy might have been converted into turbulence, and the turbulence might decay in the post-shock flow.
To maintain the relatively sharp outer boundary of the shock and of the radio relic, the spatial scale of the turbulence would have to be small,
$\la$1\arcmin\ $= 60$ kpc.
Requiring that much of the initial shock energy goes into turbulence implies that the turbulence must be transonic.

It would be useful to compare the post-shock temperature profile in Abell~3667 with other clusters with well-observed and simple radio relics and shocks.
The ``Sausage'' relic in the CIZA~J2242.8$+$5301 cluster is very narrow and has a simple curved geometry
\citep{OBR+13,OBvW+14}, while
the ``Toothbrush'' relic in the
1RXS~J0603.3$+$4214 cluster
\citep{OBvW+13} has a very sharp straight outer edge.
Of course, the shock in the Bullet cluster has a high Mach number relative to other merger shocks
\citep{MGD+02}.
However, at this time neither the Sausage or the Toothbrush have deep enough X-ray data with sufficient angular resolution to map the post-shock temperature profile.

\subsection{Particle Acceleration in the Shock}
\label{sec:discussion_accel}

Are the radio-emitting relativistic electrons in the relic being accelerated or reaccelerated in the observed shock?
We first estimate the required efficiency of this acceleration.
The flux of kinetic energy which is dissipated in the shock is given by
\begin{equation}
\Delta F_{\rm KE} = \frac{1}{2} \rho_1 v_s^3 \left( 1 - \frac{1}{C^2} \right) \, ,
\label{eq:shock_e}
\end{equation}
where $\rho_1$ is the pre-shock mass density in the gas
\citepalias{FSN+10}.
Using the values determined from the X-ray surface brightness and the shock jump conditions, we find
$\Delta F_{\rm KE} \approx 1.5 \times 10^{-4} \, {\rm erg} \, {\rm cm}^{-2} \, {\rm s}^{-1}$.
The width of the
relic from northeast to southwest is roughly 26\farcm3 or 1.63 Mpc.  Taking
the area of the shock perpendicular to the flow as a circle with this
diameter gives a perpendicular area of 2.09 Mpc$^2$.  With this size, the
total rate of conversion of shock kinetic energy is
\begin{equation}
\frac{d E_{\rm KE}}{dt} \approx 2.9 \times 10^{45} \, {\rm erg} \, {\rm s}^{-1} \, .
\label{eq:shock_dedt}
\end{equation}
In \citetalias{FSN+10}, we found that the total luminosity of radio radiation in the shock is
$L_{\rm radio} \approx 3.8 \times 10^{42}$ erg s$^{-1}$, and that including IC emission boosts
this slightly to $\sim 7 \times 10^{42}$ erg s$^{-1}$.
Since the relic has a finite width and the relativistic electrons appear to lose most of their energy within the relic (see below),
this power must be provided by the process accelerating the electrons.
This implies an efficiency of the acceleration process (the fraction of the dissipated energy which goes into acceleration) of about 0.2\%.
This is about one order of magnitude
smaller than the values of a few percent usually inferred from the radio
emission by Galactic supernova remnants \citep[e.g.,][]{RB11}.
However, this and other merger shocks have low Mach numbers compared to supernova remnant shocks,
and might be expected to be less effective at accelerating relativistic electrons
\citep[e.g.,][]{KR11,VY14}.

\subsection{Energy Losses by Relativistic Electrons}
\label{sec:discussion_loss}

The radio spectrum in the radio relic steepens with projected distance from the
outer edge of the relic \citep{RWH+97,Joh-Hol04}, consistent with a picture
in which the relativistic electrons are accelerated at the shock, and
then undergo radiative losses as they are advected away from the shock.
[Diffusion is not expected to be very important \citep[][\citetalias{FSN+10}]{BBP97}, 
and would lead to the outer edge of the radio relic being less sharp than observed if
it were significant.) 
As noted in \citetalias{FSN+10}, the loss timescale for electrons which produce
radio at a frequency
$\nu_b$ is roughly
\citep{vdLP69}
\begin{eqnarray}
t_{\rm rad} & \approx & 1.3 \times 10^8
\left( \frac{\nu_b}{1.4 \, {\rm GHz}} \right)^{-1/2}
\left( \frac{B}{3 \, \mu{\rm G}} \right)^{-3/2} \nonumber \\
& & \qquad \left[ \left( \frac{3.6 \, \mu{\rm G}}{B} \right)^2 + 1 \right]^{-1}
\, {\rm yr}
\, .
\label{eq:trad}
\end{eqnarray}
At the post-shock speed of $v_2 \approx 530$ km s$^{-1}$.  these relativistic electrons will have moved
a distance $d_{\rm rad}$ away from the shock of
\begin{eqnarray}
d_{\rm rad} & \approx & 0.07
\left( \frac{\nu_b}{1.4 \, {\rm GHz}} \right)^{-1/2}
\left( \frac{B}{3 \, \mu{\rm G}} \right)^{-3/2} \nonumber \\
& & \qquad
\left[ \left( \frac{3.6 \, \mu{\rm G}}{B} \right)^2 + 1 \right]^{-1}
\, {\rm Mpc}
\, ,
\label{eq:drad}
\end{eqnarray}
which corresponds to an angular distance of
\begin{eqnarray}
\theta_{\rm rad} & \approx & 1\farcm1
\left( \frac{\nu_b}{1.4 \, {\rm GHz}} \right)^{-1/2}
\left( \frac{B}{3 \, \mu{\rm G}} \right)^{-3/2} \nonumber \\
& & \qquad
\left[ \left( \frac{3.6 \, \mu{\rm G}}{B} \right)^2 + 1 \right]^{-1}
\cos \phi
\, ,
\label{eq:thetarad}
\end{eqnarray}
where $\phi$ is the angle between the central shock normal and the plane of
the sky.
This is consistent with the observed width of the region
at the front outer edge of the radio relic where the spectrum between 20 and
13 cm is steepens dramatically \citep{RWH+97,Joh-Hol04}. 

The fact that the full width of the relic is larger than this could indicate that
electrons are re-accelerated within the
relic, perhaps by turbulence produced by the passage of the merger shock.
Alternatively, far from the NW edge of the relic, we may be seeing radio
emission from relativistic electrons which have been recently accelerated
and which are located at the front or back edge of a convex shock and projected behind the apex of the shock. region.

\subsection{Post-Shock Electron Spectrum}
\label{sec:discussion_espect}

The observed relic radio spectrum very near to the shock is $\alpha \approx -0.7$
\citep{RWH+97,Joh-Hol04},
where the energy flux $S_\nu$ at a radio frequency $\nu$ varies as
$S_\nu \propto \nu^{- \alpha}$.
On the other hand, first order Fermi acceleration gives relativistic electrons with a power-law
spectrum $n(E) \, d E \propto E^{-p} \, d E$, where the power-law index is
$p = ( C + 2 ) / ( C - 1)$ and $C$ is the shock compression. 
The spectral
index for radio emission near the shock should be $\alpha = - ( p - 1 ) /
2 = - 3/[ 2 ( C- 1)]$
Using the value
of the compression determined above from the mean Mach number, this would give $\alpha \approx -0.87^{+0.21}_{-0.17}$,
which is marginally steeper than the observed spectrum.
If one uses the compression directly determined from the X-ray surface brightness profile $ C = 2.34^{+0.69}_{-0.36}$,
the predicted radio spectrum of the newly accelerated electrons is
$\alpha = -1.12^{+0.38}_{-0.41}$, and the difference is larger.
Similar discrepancies have been found in other clusters, and may suggest that the shock
re-accelerates a pre-existing population of relativistic electrons
\citep{KR11,POP13,SFW13}.
This would also explain why there are merger shocks without observable radio relics
in a few clusters
\citep[e.g., Abell 2146,][]{Rus+12}.
In this picture, there is no sufficient pre-existing relativistic electron population in the region of the shock
in the clusters with shocks without radio relics.
Re-acceleration of low energy relativistic electrons would also solve the problem that
low Mach number shocks are expected to be very inefficient
at accelerating thermal electrons to relativistic energies
\citep[e.g.,][]{KR11,VY14}.

\subsection{Nature of the Mushroom}
\label{sec:discussion_mushroom}

This feature consists of a tail, with a flattened region at the top and possible vortices at the two sides.
As such, it might be a rising column of buoyant, higher entropy gas.
However, the X-ray emissivity of this gas is higher than the gas around and ahead of it, indicating that
it is not buoyant (Fig.~\ref{fig:sb_mushroom}).
The temperature profile is ambiguous (Fig.~\ref{fig:mushroom_tprofile}), but is more consistent with
the Mushroom gas being cooler than the gas ahead of it, which again indicates that the specific entropy in
the Mushroom is lower than that in the surrounding gas.
Thus, it is again unlikely that the Mushroom is a buoyant feature.

The bright concentration of X-ray emission just below the top of the Mushroom coincides with the center of one of the major
subclusters of galaxies within Abell~3667 \citep[KMM2; see][]{OCN09}.
Thus, it is most likely that the Mushroom is the remnant of the cool core of this subcluster,
and that the upper edge is a cold front,
a contact interface between the subcluster cool core gas and the hotter gas in the main
cluster.

The X-ray radial surface brightness profile indicates that the X-ray emissivity jumps by a factor
of $2.75 \pm 0.25$ at the top of the Mushroom.
For the Mushroom and pre-Mushroom temperature, we will adopt the temperatures in
the two zone which are closest to the Mushroom edge, but either completely within or beyond
the edge.
This gives
$kT_1 = 6.8^{+2.4}_{-1.2}$ keV and
$kT_2 = 5.3^{+1.5}_{-0.5}$ keV,
where 2 and 1 refer to the gas within and above the Mushroom, respectively.
With these temperatures, the emissivity jump implies a jump in gas density of
a factor of 
$\rho_2 / \rho_1 = 1.63 \pm 0.08$.
It is unlikely that this is due to a shock, since the temperature within the Mushroom is,
if anything, lower than the temperature in front.
If the higher density in the Mushroom were due to a shock, it
would have a Mach number of ${\cal{M}} \approx 1.43$,
and the expected temperature jump would be also be
$T_2 / T_1 \approx 1.43$.
The observed ratio of temperatures is
$T_2 / T_1 = 0.78^{+0.26}_{-0.28}$.

Thus, it is more likely that the Mushroom is the contact interface between the cool core gas
of one of the merging subclusters, and the surrounding hot gas from the second
subcluster.
In this case, one would expect that the pressure across this contact discontinuity would
be roughly constant.
The observed ratios of the density and temperature give a pressure ratio of
$P_2 / P_1 = 1.27^{+0.43}_{-0.46}$, which is consistent with
constant pressure within the large uncertainties.

Assuming the Mushroom is a cold front, the stagnation condition at the front would allow the
Mach number and speed of the Mushroom to be determined.
The stagnation condition gives the pressure at the stagnation $P_{\rm st}$ point in terms
of the pressure upstream $P_0$ and the Mach number ${\cal{M}_{\rm cf}}$ of the flow:
\begin{equation} 
\frac{P_{\rm st}}{P_0} = \left\{
\begin{array}{ll}
\left( 1 + \frac{\gamma - 1}{2} {\cal M}_{\rm cf}^2\right)^{\frac{\gamma}{\gamma - 1}} & \\
\qquad \quad = \left( 1 + \frac{1}{3} {\cal M}_{\rm cf}^2 \right)^{5/2} \, , &
{\cal M}_{\rm cf} \le 1 \, , \\
{\cal M}_{\rm cf}^2 \,
\left( \frac{\gamma + 1}{2}
\right)^{\frac{\gamma + 1}{\gamma - 1}} \,
\left( \gamma - \frac{\gamma- 1}{2 {\cal M}_{\rm cf}^2} \,
\right)^{- \frac{1}{\gamma - 1}} & \\
\qquad \quad = \frac{4^4}{3^{5/2}} \, {\cal M}_{\rm cf}^2 \, \left( 5 - \frac{1}{{\cal M}_{\rm cf}^2} \right)^{-3/2}
\, , &
{\cal M}_{\rm cf} > 1 \, . \\
\end{array}
\right.
\label{eq:pst}
\end{equation}
Here, $\gamma$ is the adiabatic index, and the second expressions apply for $\gamma = 5/3$.
A Mach number of 1 corresponds to the pressure ratio of
$P_{\rm st} / P_0 = (4/3)^{5/2} \approx 2.0528$.
Given a measured value of $P_{\rm st} / P_0$, equation~(\ref{eq:pst}) can be inverted to give
the Mach number.
For subsonic flow, this is simply
\begin{equation}
{\cal M}_{\rm cf}^2 = 3 \left[ \left( \frac{P_{\rm st}}{P_0} \right)^{2/5} - 1 \right] \qquad  \frac{P_{\rm st}}{P_0} \le (4/3)^{5/2}
\, .
\label{eq:mst_1}
\end{equation}
For supersonic flow, it is easiest to iterate the solution using
\begin{equation}
\left( {\cal M}_{\rm cf}^2 \right)_{i+1}
= \frac{3^{5/2}}{4^4} \, \left[ 5 - \frac{1}{\left( {\cal M}_{\rm cf}^2 \right)_i} \right]^{3/2} \, \left( \frac{P_{\rm st}}{P_0} \right)
\qquad \frac{P_{\rm st}}{P_0} > (4/3)^{5/2}
\, .
\label{eq:mst_2}
\end{equation}
Here, $\left( {\cal M}_{\rm cf}^2 \right)_{i+1}$ is the improved estimate based on the previous estimate $\left( {\cal M}_{\rm cf}^2 \right)_{i}$.
This converges rapidly if one starts with an estimate which is greater than unity, say $\left( {\cal M}_{\rm cf}^2 \right)_0 = 2$.

The difficulty in applying this method is that $P_0$ needs to be the initial pressure of the gas currently at the stagnation point.
Because of the radial pressure gradient in a cluster, $P_0$ cannot be assumed to be the pressure well ahead of the cold front.
We evaluated the pressure profile in the regions from 132\arcsec\ to 272\arcsec\ ahead of the Mushroom, and extrapolated this to the position of the
Mushroom.
This was compared the the pressure at the top of the Mushroom.
This gave $P_{\rm st} / P_0 = 2.05^{+0.60}_{-0.27}$, which is very close to the critical transonic pressure ratio.
The implied Mach number is ${\cal M}_{\rm cf} = 1.0 \pm 0.1$, and the speed of the Mushroom
relative to the gas ahead of it would be
$1340^{+250}_{-140}$ km s$^{-1}$.
However, due to the assumptions involved is this estimate, and the fact the the region ahead of the Mushroom includes the lower
portion of the radio relic and eventually the observed merger shock, the systematic uncertainties are much larger than
the statistic ones.
Thus, the motion of the Mushroom relative to the gas ahead of it could either be subsonic or
supersonic.

The radial temperature profile along the Mushroom (Fig.~\ref{fig:mushroom_tprofile})
shows a possible temperature drop roughly 130\arcsec\ above the top of the Mushroom.
This could indicate that there is a shock at this location, although the temperature errors
are very large.
On the other hand, the X-ray surface brightness profile has no significant feature at this
location, and is inconsistent with the emissivity jump expected from a shock which would
give the temperature jump.

Alternatively, could the Mushroom be the ``piston'' which has driven the merger shock at the
top of the radio relic?
As measured from the center of the cluster, the apex of the radio relic shock is at a
position angle (measured to the east from north) of about $-33\arcdeg$,
while the center of the Mushroom is at a position angle of
about $-54\arcdeg$.
While this $\sim$20\arcdeg\ difference might indicate they have different origins,
it could also be the result of an off-axis merger, or an elongated distribution in the original
primary cluster.

For a solid object moving at a constant supersonic speed in steady-state through a uniform density medium,
the offset between the object and the bow shock depends on the Mach number of the motion
and the radius of curvature of the object.
Based on the elliptical fit to the shape of the top of the Mushroom  (Fig.~\ref{fig:sb_mushroom_regions}),
the radius of curvature of the Mushroom is about $R_{\it cf} \approx 10\farcm8$, while the difference in the radius
of the apex of the shock from that of the Mushroom is about $d_s \approx 9\farcm5$.
This gives a ratio of $d_s / R_{\it cf} \approx 0.88$.
For a shock Mach number of 
${\cal{M}} = 2.54^{+0.80}_{-0.43}$,
the predicted shock offset for a spherical object is $d_s / R_{\it cf} \approx 0.32^{+0.08}_{-0.06}$
\citep{Sch82,VMM01b,Sar02}.
While this is much smaller that the observed separation, the same is true for most of the observed shock/cold-front pairs in
clusters
\citep[e.g.,][]{Das+16b}.
At a time after first core passage, the cool core of the merging subcluster is being slowed in its motion by both gravity and
ram pressure, while the shock is accelerating down the declining pressure and density gradient
in the outer cluster.
Thus, one expects the shock to move out well ahead of the cold front, as is confirmed in numerical simulations
\citep[e.g.,][]{MCO16}.

The shock velocity of 
$v_s = 1450^{+460}_{-260}$ km s$^{-1}$ is not much larger than the very uncertain
value derived for the cold front at the Mushroom of
$v_{\rm cf} = 1340^{+250}_{-140}$ km s$^{-1}$.
However, the later velocity is relative to the post-shock gas.
For a simple plane parallel shock, the post-shock velocity would be
$v_2 = v_s / C = 530^{+60}_{-50}$ km s$^{-1}$.
Then, the speed of the shock relative to the cold front in a consistent frame would
be $ v_s - v_s / C - v_{\rm cf} = -420^{+340}_{-310}$ km s$^{-1}$.
A negative value is probably not consistent with the Mushroom being the piston behind the
radio relic shock, at least at the observed stage of the merger.
However, the statistical errors are large, and the systematic errors due to the simplified assumptions
made in this analysis are probably even larger.
Numerical simulations would be useful to test if the observed geometry and velocities in
the northern region of Abell 3667 could be the result of a simple binary offset merger.

Another possibility might be that the SW portion of the Notch on the radio relic is actually
due to a separate shock driven by the Mushroom.
However, neither the X-ray surface brightness profile
(Fig.~\ref{fig:sb_mushroom}) nor the temperature profile
(Fig.~\ref{fig:mushroom_tprofile}) show any evidence for
a shock at this location (a radius of $\approx 1700\arcsec$).

On the other hand, the top of the Mushroom coincides with the
sharp SW bottom edge of the notch in the NW radio relic.
This is consistent with the Mushroom being the cool core of
a subcluster which merged from the SE, and the
NW radio relic being due to a shock driven into the subcluster
which merged from the NW.
The top of the Mushroom is a contact interface between gases having different
origins.
Relativistic electrons accelerated by the shock in the gas from the NW subcluster
would not enter into the Mushroom, unless there were significant diffusion or mixing
at its upper surface.
The diffusion lengths are expected to be fairly short
\citepalias{FSN+10}.
The relatively sharp appearance of the upper edge of the Mushroom
suggests that this surface is not unstable, and that there hasn't been
much mixing there.
Then, one would not expect that relativistic electrons would have
passed into the Mushroom.
Some relic radio emission seen in projection against the Mushroom might be
produced by particles in the NW merging subcluster which were advected around the Mushroom.

\section{Conclusion}
\label{sec:conclusion}

We have analyzed a long series of \xmmnewton\ observations of the NW radio relic region of the Abell 3667
cluster of galaxies, as well as the previous \xmm\ observations of the cluster.
We detect a discontinuity in the slope of the X-ray surface brightness at
the sharp outer edge of the radio relic, which is well-fit by a model for
a discontinuity in the X-ray emissivity by a factor of
$R = 3.11^{+1.75}_{-0.94}$.
At the same location, we detect a jump in the temperature of the gas
by a factor of
$( T_2 / T_1 ) = 4.35^{+1.91}_{-1.05}$.

Both of these are consistent with a merger shock being located at the outer edge of the NW radio relic.
The jump in X-ray emissivity from the surface brightness profile implies a shock
compression of
$ C \equiv \rho_2 / \rho_1 = 2.34^{+0.69}_{-0.36}$,
leading to a
shock Mach number of ${\cal{M}} = 2.05^{+0.73}_{-0.38}$.
The temperature jump implies a Mach number of
${\cal{M}} = 3.34^{+0.91}_{-0.50}$.
The Mach number from the temperature jump is 1.46 $\sigma$ higher than the value
from the X-ray surface brightness jump.
It is not clear whether this represents a departure from the Rankine--Hugoniot jump conditions
for a perpendicular shock in a $\gamma = 5.3$ non-magnetized perfect gas,
or if it is just a large statistical fluctuation or the result of systematic errors or projection.

The mean shock Mach number from the surface brightness profile and temperature jump is
${\cal{M}} = 2.54^{+0.80}_{-0.43}$.
Although the uncertainties are large, this Mach number makes the NW Relic Shock in Abell 3667
one of the strongest shocks seen in a cluster,
although weaker than that in the Bullet Cluster
\citep[${\cal{M}} = 3.0 \pm 0.4$;][]{MV07}.
This Mach number implies a shock compression of
$C = 2.73^{+0.55}_{-0.29}$,
and the shock speed is
$v_s = {\cal{M}} c_s = 1450^{+460}_{-260}$ km s$^{-1}$.

The X-ray surface brightness profile at the shock is well-fit by a sharp discontinuity in
the gas density.
However, the gas temperature rises more gradually within the shock region over a scale of
several arc minutes, which is much broader than might have been expected based on the
 \xmmnewton\ PSF.
Projection accounts for part of this gradual rise, but not most of it;
the de-projected temperature profile still requires about 2\farcm3 to rise from the
pre-shock value to a plateau at the post-shock value.
We considered a non-equipartition model in which electrons are only heated
adiabatically at the shock, and
the electron temperature increases gradually as a result of Coulomb
collisions with ions.
The pre- and post-shock densities and velocities are well-determined by
the X-ray observations and the shock jump conditions.
However, the electron temperature still increases more rapidly behind the
shock than is observed.
We also considered if the slow rise in the gas temperature was
due to non-equilibrium ionization, but this cannot explain the gradual temperature
increase.

It is possible that the gradual temperature rise at the shock is the result of inhomogeneities in
the gas at the shock front.
It is also possible that, if the simple, self-similar ellipsoidal X-ray emissivity model we have assumed is incorrect,
this might also affect the observed temperature profile behind the shock.
One way to test this methodology would be to simulate the X-ray images and spectra of shocks in detailed, high-resolution numerical simulations of merging clusters.
One could analyze the simulated X-ray properties of shocks in these clusters from a variety of viewing angles,
and use the simple analytic model discussed here to determine the parameters of the shocks.
These could be compared to the actual shock properties in the numerical simulations.

Another interesting possibility to explain the slow rise in the post-shock temperature is that the shock energy is initially dissipated into some mix of thermal and nonthermal energy, and the nonthermal energy decays into thermal energy in the post-shock region.
For example, part of the shock energy might have been converted into turbulence, and the turbulence might decay in the post-shock flow.
Requiring that much of the initial shock energy goes into turbulence implies that the turbulence must be transonic.

At present, accurate temperature profiles of cluster merger shocks are rare.
It would be useful to obtain deeper X-ray data on several strong, well-defined cluster
merger shocks
to see if this gradual temperature rise is a general phenomena, which is telling us
something about the physics of merger shocks.
Good examples might include the ``Sausage'' relic in the CIZA~J2242.8$+$530
\citep{OBR+13,OBvW+14} and
the ``Toothbrush'' relic in the
1RXS~J0603.3$+$4214 cluster
\citep{OBvW+13}.

The presence of a merger shock at the sharp outer edge of
the NW radio relic in Abell 3667 is consistent with the
prediction of theories in which the relativistic electrons in radio relics
are accelerated or re-accelerated by the shock.
Comparing the required energy input in relativistic electrons in the NW radio relic
with the kinetic energy dissipated in the observed shock,
we find that the efficiency of acceleration of electrons in the shock
is about 0.2\%.
This is lower than the values of a few percent usually inferred from the radio
emission by Galactic supernova remnants \citep[e.g.,][]{RB11}.
However, this and other merger shocks have low Mach numbers compared to supernova remnant shocks,
and might be expected to be less effective at accelerating relativistic electrons
\citep[e.g.,][]{KR11,VY14}.

The radio spectrum in the radio relic steepens with projected distance from the
outer edge of the relic \citep{RWH+97,Joh-Hol04}, consistent with a picture
in which the relativistic electrons are accelerated at the shock, and
then undergo radiative losses as they are advected away from the shock.
We found that the length scale over which the radio spectrum steepens is
consistent with the expected timescale of radiative losses by the
radio emitting electrons, and the post-shock speed determined from
the X-ray observations and the shock jump conditions.
The full width of the relic is larger than the length scale associated with
radiative losses.
A simple explanation of this is that, far from the NW edge of the relic, we may be seeing radio
emission from relativistic electrons which have been recently accelerated
and which are located at the front or back edge of a convex shock region.
Alternatively,
the width of the relic may indicate that
electrons are re-accelerated within the
relic, perhaps by turbulence produced by the passage of the merger shock.

While these results would all be consistent with a model in
which the radio electrons are directly accelerated from the thermal
distribution by first-order Fermi diffusive shock acceleration,
the radio spectrum immediately behind the shock front is not.
The observed spectrum is flatter than predicted by diffusive shock acceleration
theory given the observed shock compression.
Similar discrepancies have been found in other clusters, and may suggest that the shock
re-accelerates a pre-existing population of relativistic electrons
\citep{KR11,POP13,SFW13}.
This would also solve the problem that low Mach number shocks are expected to be very inefficient
at accelerating thermal electrons to relativistic energies
\citep[e.g.,][]{KR11,VY14}.

The other new feature identified in our \xmm\ observation is the Mushroom, which
consists of a tail, with a flattened region at the top and possible
vortices at the two sides.
There is a bright region of X-ray emission just below the top of the Mushroom which
coincides with the center of one of the major
subclusters of galaxies within Abell~3667 \citep[KMM2; see][]{OCN09}.
The temperature and entropy of the Mushroom are smaller than
the gas ahead (to the north) of it, and the pressure is
nearly continuous.
This indicates that the Mushroom is most likely a cold front containing
the remnant cool core of a merging subcluster.

We propose that the NW radio relic shock is a merger shock driving by the subcluster associated
with the Mushroom.
In this scenario, most of the features of Abell 3667 would be the result of a binary off-axis merger
of two subclusters, and the observed epoch is well after first core passage.
The relic shock is at a radius of 2.13 Mpc from the cluster center.
Given the current shock velocity, this corresponds to a
propagation time of about $(1.4 / \cos i)$ Gyr, where
$i$ is roughly the angle between the merger axis and the plane of
the sky.
This should give a very crude estimate of the time since first core passage.
The subcluster corresponding to the Mushroom is probably the less massive one
and collided from the SE.
The famous cold front near the center of Abell 3667
\citep{VMM01b}
would be the remnant cool core of the more massive merging subcluster, which collided from
the NW.
In this scenario,
the fainter, smaller SE radio relic in Abell 3667 is due to acceleration at a merger shock
which was driven into the smaller subcluster by the core of the more massive subcluster.

This model predicts that sufficiently deep X-ray or SZ observations should detect a merger
shock at the outer (SE) edge of the SE radio relic.
However, detailed numerical simulations would be very useful to test the consistency of
this simple merger scenario.
While the early and pivotal computational model of \citet{RBS99} and
the more recent simulations from  \citet{DSB+14} are broadly consistent
with this picture, the latter calculations are were not specifically tuned to this
cluster, and the earlier simulations could not make use of all of the
very detailed new X-ray and radio observations of Abell 3667.

\acknowledgments
CLS thanks the Institute for Astro- and Particle Physics at the University of Innsbruck and the
Eramus Mundus Program of the European Commission for their hospitality and
support for the period when this project was started.
He also thanks the Physics Department at the University of Helsinki for their hospitality
as the work was being completed.
This research was primarily supported
by NASA ADAP grant NNX11AD15G, but also by
NASA {\it XMM} grant NNX15AG26G and {\it Chandra} grants
GO1-12169X, GO4-15123X, and GO5-16131X.

\appendix
\section{A.\quad Elliptical Pie Annuli Counts for an Ellipsoidal Emissivity Jump}
\label{sec:ellipt}

Here, we present the expressions used to accumulate the X-ray surface brightness profile of the radio relic regions
(\S~\ref{sec:sb}).
Our treatment follows that in
\citet[][see Appendices A.1 and A.2]{Kor+11},
which for a single emissivity edge is formally equivalent to that given in
\citet{VMM01b}, although our use of the incomplete beta function is much more computationally efficient than the
use of the hypergeometric function in the earlier paper.
The basic assumption is that the X-ray emissivity $\epsilon$ is constant on ellipsoidal surfaces, so
that $\epsilon = \epsilon ( r )$, where $r \equiv [ ( x / a )^2 + ( y / b)^2 + ( z / c )^2 ]^{1/2}$ is the scaled ellipsoidal radius.
The center of the ellipsoid is the point ($x=0$, $y=0$, $z=0$), and the three cartesian coordinates $(x,y,z)$ are taken to be parallel to the principle axes of the ellipsoid.
We assume that the emissivity has a jump at $r = 1$, and is a power-law on either side of this jump
(eqn.~\ref{eq:emiss}),
Thus, $a$, $b$, $c$ are the semi-axes of the ellipsoidal shape of the emissivity jump.
We assume the $x$-axis (semi-axis $a$) and $y$-axis (semi-axis $b$) lie in the plane of the sky, and that the
$z$-axis lies along our line-of-sight.

The projected X-ray surface brightness will then be a function of the projected elliptical radius,
\begin{equation}
A(x,y) \equiv  \left( \frac{x^{2}}{a^{2}}+\frac{y^{2}}{b^{2}} \right)^{1/2} \, .
\label{eq:A}
\end{equation}
Combining equations~(A3) and (A6) in Appendix A.1 of \citet{Kor+11}
gives 
equation~(\ref{eq:sbi}),
while combining
equations~(A3) and (A8) gives
equation~(\ref{eq:sbo}).
The normalizations of the surface brightness profile are given by
\begin{equation}
I_i = \frac{1}{4 \pi (1 + z_r)^2} \, \epsilon_{i} \, c \, B( p_i - \frac{1}{2}, \frac{1}{2} )
\, ,
\label{eq:Ii}
\end{equation}
and
\begin{equation}
I_o = \frac{1}{4 \pi (1 + z_r)^2} \, \epsilon_{o} \, c \, B( p_o - \frac{1}{2}, \frac{1}{2} )
\, ,
\label{eq:Io}
\end{equation}
where $z_r$ is the cluster redshift, and $B(u,v)$ is the beta function.
Here, we assume the surface brightness is given in count units (e.g., cts s$^{-1}$ arcsec$^{-2}$);
that is, it is multiplied by the instrument collecting area and integrated over the observed passband.
The factor of $(1 + z_r)^2$ differs from that given in \citet{Kor+11} because in that paper the surface brightness
was not averaged over the instrument response.
The ratio of equations (\ref{eq:Ii}) and (\ref{eq:Io}) leads immediately to equation~(\ref{eq:ii}).

Typically, the X-ray surface brightness profile will be determined from counts accumulated in elliptical pie annuli
(``epandas''), where the ellipse is similar in shape to the projected edge of the X-ray emissivity.
Assume that one such epanda extends from $A_1$ to $A_2$ radially, and between the angles $\phi_1$ and $\phi_2$ azimuthally, where $\phi$ is measured from the $x$-axis (semi-major axis $a$).
Assume that $a$ and $b$ are measured in the same angular units as used for the surface brightness (e.g., arcsec).
Then, the counts in this epanda will be given by
\begin{equation}
C = C_{\rm in} + C_{\rm out} \, ,
\label{eq:C}
\end{equation}
where
\begin{equation}
C_{\rm in} = D I^\prime_i \left[ F_{\rm in} ( A_2 ) - F_{\rm in} ( A_1 ) \right] \, ,
\label{eq:Cin}
\end{equation}
and
\begin{equation}
C_{\rm out} = D I_o \left[ F_{\rm out} ( A_2 ) - F_{\rm out} ( A_1 ) \right] \, .
\label{eq:Cout}
\end{equation}
Here, $D$ gives the area of an elliptical pie wedge, so that
\begin{equation}
D = \frac{ a b }{2} \left[ \tan^{-1} \left( \frac{ \tan \phi }{b/a} \right) \right]^{\phi_2}_{\phi_1}
\, .
\label{eq:D}
\end{equation}
Note that for a full elliptical annulus, $\phi_2 = 180\arcdeg = \pi$ radians and
$\phi_1 = -180\arcdeg = -\pi$ radians, and
$D = \pi a b$ as expected.

The value of $I^\prime_i$ is given by
\begin{equation}
I^\prime_i = \frac{ I_i }{ B( p_i - \frac{1}{2}, \frac{1}{2} ) } = \frac{ R \, I_o }{ B( p_o -  \frac{1}{2} ,  \frac{1}{2} )}
= 
\frac{1}{4 \pi (1 + z_r)^2} \, \epsilon_{i} \, c
\, .
\end {equation}
The purpose of this definition is to avoid the problems with the definition of $I_i$ which occur when
$B( p_o - \frac{1}{2} , \frac{1}{2} )$ or $B( p_i - \frac{1}{2} , \frac{1}{2} )$ are zero or infinity ($ p_o, p_i = \frac{1}{2}$, 0, $-\frac{1}{2}$, $-1$, $-\frac{3}{2}$, \dots).

The values of $F_{\rm in}$ and $F_{\rm out}$ are determined by integrating the surface brightness over the areas of the elliptical pie wedges.
Since a power-law emissivity can diverge at either large or small radii, the values are accumulated from the projected emissivity edge, $A = 1$.
With this definition, the integrals are given by
\begin{equation}
F_{\rm in} (A) = 
\left\{
\begin{array}{cc}
\frac{1}{\frac{3}{2} - p_i } \left[
A^{-2 p_i + 3} B_{1 - A^2} \left( \frac{1}{2} , p_i - \frac{1}{2} \right) - 2 \sqrt{1 - A^2} \right] & 0 < A^2 < 1 \\
0   & A^2 \ge 1
\end{array}
\right.
\, ,
\label{eq:Fin}
\end{equation}
for $ p_i \ne \frac{3}{2}$.
Here, $B_x ( u , v )$ is the unnormalized incomplete beta function.
Note that $F_{\rm in} (A)$ is negative for $A^2 < 1$ and is always zero for $A^2 \ge 1$.
This expression is regular for all other values of $p_i$;
however, some algorithms for computing the unnormalized incomplete beta function
will fail for
for $ p_i = \frac{1}{2}$, $-\frac{1}{2}$, $-1$, $-\frac{3}{2}$, \dots.
For the special case $ p_i = \frac{3}{2}$, 
the expression is
\begin{equation}
F_{\rm in} (A) = 
\left\{
\begin{array}{cc}
4 \sqrt{1 - A^2}  - 2 \ln \left( 1 + \sqrt{1 - A^2} \right) + 2 \ln \left( 1 - \sqrt{1 - A^2} \right) & 0 < A^2 < 1 \\
0   & A^2 \ge 1
\end{array}
\right.
\, .
\label{eq:Finspec}
\end{equation}

For the accumulated counts from the outer emission region, the expression is
\begin{equation}
F_{\rm out} (A) = \frac{1} {\frac{3}{2} - p_o }  
\left\{
\begin{array}{cc}
A^{-2 p_o + 3} I_{A^2} \left( p_o - \frac{1}{2} , \frac{1}{2} \right) + 2 \frac{\sqrt{1 - A^2}}{B( p_o - \frac{1}{2} , \frac{1}{2} )} -1 & 0 < A^2 < 1 \\
A^{-2 p_o + 3} - 1 & A^2 \ge 1
\end{array}
\right.
\, .
\label{eq:Fout}
\end{equation}
The integrals diverge at large radii for $ p_o < 1/2$, so we restrict the fits to $ p_o > 1/2$.
Again, the value $p_o = 3/2$ requires special handling.
For $p_o = 3/2$, the expression is
\begin{equation}
F_{\rm out} (A) =
\left\{
\begin{array}{cc}
\ln \left( A^2 \right) - 2 \sqrt{1 - A^2}
+ \ln \left( 1 + \sqrt{1 - A^2} \right) - \ln \left( 1 - \sqrt{1 - A^2} \right) & 0 < A^2 < 1 \\
\ln \left( A^2 \right) & A^2 \ge 1
\end{array}
\right.
\, .
\label{eq:Foutspec}
\end{equation}

\section{B.\quad Deriving Shock Jump Conditions}
\label{sec:shockjump}

Here, we describe the process of determining the shock Mach number ${\cal{M}}$ from
X-ray measurement of the emissivity (Appendix~\ref{sec:ellipt}) or temperature jump at a shock, or Sunyaev-Zel'dovich measurements of the pressure jump
\citep[e.g.,][]{Kor+11}.
We consider a shock in a gas with an adiabatic index of $\gamma$, but also give the simpler expressions
for the most useful case of $\gamma = 5/3$.
The observed signatures of shocks all arise from the standard
Rankine--Hugoniot jump conditions
\citep[e.g.,][]{LL59}.

\subsection{B.1.\quad Shock X-ray Surface Brightness Discontinuity}
\label{sec:shocksb}

X-ray surface brightness profiles give the jump in the X-ray emissivity $R$ with a particular instrument and in a certain energy band
(Appendix~\ref{sec:ellipt}).
The X-ray emissivity can be written as
$\epsilon = \Lambda(T, Z) n_e^2$, where $\Lambda(T, Z)$ gives the dependence of the emission on the electron temperature $T$ and composition $Z$.
The shock compression is $C \equiv \rho_2 / \rho_1 = n_2 / n_1$, where $n_1$ ($n_2$) are the pre-shock (post-shock) electron density, $\rho$ is the mass density, and the subscripts 1 and 2 refer to pre- and post-shock gas, respectively.
Then, the observed ratio of emissivities is
$ R = \epsilon_2 / \epsilon_1 = ( \Lambda_2 n_2^2 ) / ( \Lambda_1 n_1^2 ) = ( \Lambda_2 / \Lambda_1 ) C^2$.
Thus, the shock compression is given by
\begin{equation}
C = R^{1/2} \left( \frac{\Lambda_2}{\Lambda_1} \right)^{-1/2} 
\, .
\label{eq:compress}
\end{equation}

The ratio of emissivity functions $ ( \Lambda_2 / \Lambda_1 )$ can be derived from the spectral fits for the pre-and post-shock regions used to derive the temperatures.
One reads each of the best-fit spectral models into the fitting software (e.g., {\sc xspec}%
\footnotemark[7]),
and removes any foreground/background components, leaving only the cluster thermal gas emission.
The instrument responses and energy passband should be those for the instrument and energy range used to construct the X-ray surface brightness.
Then, one sets the normalization of the only remaining thermal component to a fixed value (say unity, or a smaller fixed value if pile-up is included in the fits).
One determines the model count rate in the observed band, CR$_1$ for the pre-shock model and CR$_2$ for the post-shock model.
Then,
\begin{equation}
\left( \frac{\Lambda_2}{\Lambda_1} \right) = \left( \frac{{\rm CR}_2}{{\rm CR}_1} \right)
\, .
\label{eq:emissfunc}
\end{equation}
The errors in the temperatures (and possibly abundances) for the spectral fits can be used to determine the
errors in CR$_1$ and CR$_2$, and thus the error $\delta ( \Lambda_2 / \Lambda_1 )$ in the emission function ratio.
Then, the error in the compression ratio is
\begin{equation}
\delta C = \frac{1}{2} \, C
\left\{ \left[ \frac{\delta R}{R} \right]^2
+ \left[ \frac{\delta ( \Lambda_2 / \Lambda_1 )}{( \Lambda_2 / \Lambda_1 )} \right]^2 \right\}^{1/2}
\, .
\label{eq:comperror}
\end{equation}

Given the shock compression, the square of the Mach number is
\begin{equation}
{\cal{M}}^2 = \frac{2 C}{( \gamma + 1 ) - ( \gamma - 1 ) C } = \frac{3 C}{4 - C}
\, ,
\label{eq:m2compress}
\end{equation}
where the rightmost expression here and below refers to $\gamma = 5/3$.
The error in the square of the Mach number is given by
\begin{equation}
\left( \delta {\cal{M}}^2 \right)
=
\frac{ 2 ( \gamma + 1 ) }{ \left[ ( \gamma + 1 ) - ( \gamma - 1 ) C \right]^2 }
\, \delta C
=
\frac{ 12 }{ ( 4 - C )^2 } \, \delta C
\, .
\label{eq:m2comperr}
\end{equation}
Of course,
\begin{equation}
\delta {\cal{M}}
=
\frac{1}{2} \, \frac{ \left( \delta {\cal{M}}^2 \right) }{ {\cal{M}} }
\, .
\label{eq:merr}
\end{equation}

\subsection{B.2.\quad Shock X-ray Temperature Jump}
\label{sec:shocktjump}

X-ray spectral of regions projected in front of and behind the shock position can be used to estimate the pre-
and post-shock temperatures, $T_1$ and $T_2$ respectively.
The temperature jump condition for a shock is
\begin{equation}
\frac{T_2}{T_1} = \frac{ \left[ 2 \gamma  {\cal M}^2 - ( \gamma - 1 ) \right]
\left[  ( \gamma - 1 ) {\cal M}^2 + 2 \right]}{( \gamma + 1 )^2  {\cal M}^2}
= \frac{5 {\cal{M}}^4 + 14 {\cal{M}}^2 - 3}{16 {\cal{M}}^2} 
\, .
\label{eq:tjump}
\end{equation}
The temperature jump condition is a bit more complicated than that for the density or pressure because temperature in an intensive quantity, and the standard
Rankine--Hugoniot jump conditions
\citep[e.g.,][]{LL59} conserve the mass, momentum, and energy fluxes.
Define
\begin{equation}
\psi \equiv ( \gamma + 1 )^2 \left( \frac{T_2}{T_1} + 1 \right) - 8 \gamma
=
\frac{8}{9} \, \left( 8 \frac{T_2}{T_1} - 7 \right)
\, .
\label{eq:shockpsi}
\end{equation}
Then,
\begin{equation}
{\cal{M}}^2 = \frac{ \psi + \sqrt{ \psi^2 + 16 \gamma ( \gamma - 1 ) ^2 }}{4 \gamma ( \gamma - 1 )}
= \frac{ \left( 8 \frac{T_2}{T_1} - 7 \right) + \left[ \left( 8 \frac{T_2}{T_1} - 7 \right)^2 + 15 \right]^{1/2} }{5}
\, ,
\label{eq:m2temp}
\end{equation}
where the far right hand side assumes $\gamma = 5/3$.

The error in the shock temperature jump is given by
\begin{equation}
\left[ \frac{ \delta ( T_2 / T_1 ) }{ ( T_2 / T_1 ) } \right]^2
=
\left( \frac{ \delta T_1 }{ T_1 } \right)^2
+
\left( \frac{ \delta T_2 }{ T_2 } \right)^2
- 2\,  \frac{\sigma(T_1,T_2)}{T_1 T_2 }
\, . 
\label{eq:temperr}
\end{equation}
Here, $\sigma(T_1,T_2)$ is the covariance of the two temperatures.
If the observed projected temperatures are used, then they are independent and
$\sigma(T_1,T_2) \equiv 0$.
However, in this case for most geometries, the value of $T_2$ will be underestimated due to cooler gas
at temperature $T_1$ projected on the same region.
If the temperatures are de-projected, then the errors will be correlated,
and $\sigma(T_1,T_2) \ne 0$.
[Note that standard spectral fitting programs give covariances for de-projected temperatures
(e.g., {\sc xspec}%
\footnotemark[7]).]

The error in the square of the Mach number is
\begin{equation}
\left( \delta {\cal{M}}^2 \right)
=
\frac{ ( \gamma + 1 )^2 {\cal{M}}^2 }{ \sqrt{ \psi^2 + 16 \gamma ( \gamma - 1 ) ^2 } } \, \delta ( T_2 / T_1 )
=
\frac{ 8 {\cal{M}}^2 }{ \left[ \left( 8 \frac{T_2}{T_1} - 7 \right)^2 + 15 \right]^{1/2} }
\, \delta ( T_2 / T_1 )
\, ,
\label{eq:m2temperr}
\end{equation}
and the error in the Mach number is given by equation~(\ref{eq:merr}).

\subsection{B.3.\quad Shock Sunyaev-Zel'dovich Effect Pressure Jump}
\label{sec:shockpjump}

High spatial resolution observations of the Sunyaev-Zel'dovich effect give a map of the Compton $y$ parameter,
which is proportional to the integral of the electron pressure through the cluster.
Thus, these images can be analyzed using the same techniques as X-ray images (\S~\ref{sec:ellipt})
to give the jump in electron pressure at a shock
\citep[e.g.,][]{Kor+11}.
Assuming electron-ion equipartition, this is equal to the jump in total gas pressure at the shock.
From the shock pressure jump condition,
the square of the Mach number is
\begin{equation}
{\cal{M}}^2 = \frac{ ( \gamma + 1 ) \frac{P_2}{P_1} + ( \gamma - 1 )}{2 \gamma} =
\frac{4 \frac{P_2}{P_1} + 1 }{5} \, .
\label{eq:m2pjump}
\end{equation}
The error in the square of the Mach number is then
\begin{equation}
\left( \delta {\cal{M}}^2 \right)
=
\frac{( \gamma + 1 )}{2 \gamma} \, \delta ( P_2 / P_1 )
=
\frac{4}{5} \, \delta ( P_2 / P_1 )
\, ,
\label{eq:m2presserr}
\end{equation}
and the error in the Mach number is given by equation~(\ref{eq:merr}).

\subsection{B.4.\quad Deriving the Adiabatic Index Given the Shock Compression and Temperature Jump}
\label{sec:gamma}

The X-ray surface brightness profile and temperature profile at a shock give two independent measures of the
shock Mach number.
In principle, these can be used to derive both the Mach number ${\cal{M}}$ of the shock and the adiabatic index $\gamma$ of the gas.
Here, we assume that the shock compression $C \equiv \rho_2 / \rho_1$ (eq.~\ref{eq:compress}) and temperature jump
$T_2 / T_1$ (eq.~\ref{eq:tjump}) have been determined.
Then, requiring that they be consistent with the same Mach number for some value of $\gamma$ gives
\begin{equation}
\gamma =
\left[ \frac{  ( P_2 / P_1 ) - 1 }{ ( P_2 / P_1 ) + 1} \right]  \, \left( \frac{ C +1 }{ C -1 } \right)
=
\left[  \frac{ C  (T_2 / T_1 ) - 1 }{ C ( T_2 / T_1 ) + 1 } \right] \, \left( \frac{ C +1 }{ C -1 } \right)
\, .
\label{eq:gamma}
\end{equation}
The Mach number is then given by
any of equations~(\ref{eq:m2compress}), (\ref{eq:m2temp}), or (\ref{eq:m2pjump}).
The errors can be propagated using the derivatives
\begin{equation}
\frac{\partial \gamma}{\partial C} =
2 \gamma \left[
\frac{ T_2 / T_1 }{ \left( C T_2 / T_1 \right)^2 - 1 }
-
\frac{1}{ C^2 - 1 }
\right]
\, ,
\label{eq:dgdc}
\end{equation}
\begin{equation}
\frac{\partial \gamma}{\partial ( T_2 / T_1 ) } =
\frac{ 2 \gamma C }{ \left( C T_2 / T_1 \right)^2 - 1 }
\, ,
\label{eq:dgdt}
\end{equation}
\begin{equation}
\frac{\partial {\cal{M}}^2}{\partial C} =
\frac{{\cal{M}}^2}{C} + \frac{{\cal{M}}^4}{2 C}
\left[ ( \gamma - 1 ) + ( C - 1 ) \, \frac{\partial \gamma}{\partial C} \right]
\, ,
\label{eq:dm2dc}
\end{equation}
\begin{equation}
\frac{\partial {\cal{M}}^2}{\partial (T_2 / T_1 )} =
\frac{{\cal{M}}^4 ( C - 1 ) }{2 C} \,
\frac{\partial \gamma}{\partial ( T_2 / T_1 )}
\, .
\label{eq:dm2dt}
\end{equation}

\subsection{B.5.\quad Shocks with a Parallel Magnetic Field}
\label{sec:magnetic}

Consider a plane-parallel shock with a uniform magnetic field which is parallel to the shock front.
Let $B_1$ ($B_2$) be the pre-shock (post-shock) field.
Assume that the plasma has a high conductivity, so that the field is frozen into the plasma.
Then, the shock jump conditions
\citep[][eqs.~(7)--(11)]{Hel53}
are identical to those for an unmagnetized gas, except for the addition of the magnetic pressure
$B^2/(8 \pi)$ to the momentum equation and the enthalpy flux $vB^2/(4 \pi)$ to the energy equation.
The frozen-in condition implies that the pre- and post-shock fields are related by
$B_2 / \rho_2 = B_1 / \rho_1$.
It is useful to characterize the shock by the Mach number ${\cal{M}}$ (for consistency with the unmagnetized
shock models) and by the ratio of the magnetic to gas pressure in the pre-shock region
$ b \equiv B_1^2 / ( 8 \pi P_1 )$.
As in the cases above, X-ray observations can be used to determine the shock compression
$C = \rho_2 / \rho_1$
(\S~\ref{sec:shocksb})
and the temperature jump $T_2 / T_1$
(\S~\ref{sec:shocktjump}).
These can be used to determine the shock pressure jump $ p = P_2 / P_1 = C ( T_2 / T_1 )$, or this may have been derived from SZ measurements
(\S~\ref{sec:shockpjump}).
Then, the shock jump conditions can be solved (after considerable algebra) for the pre-shock ratio of
magnetic to gas pressure
\begin{equation}
b \equiv \frac{B_1^2}{8 \pi P_1} =
\frac{2 ( p - 1 ) - ( C - 1 )[ 2 \gamma + ( \gamma - 1 ) ( p - 1 )]}
{( \gamma - 1 ) ( C - 1 )^3}
=
\frac{3 ( p - 1 ) - ( C - 1 )( p + 4 ) }
{( C - 1 )^3}
\, ,
\label{eq:magb}
\end{equation}
while the Mach number is given by
\begin{equation}
{\cal{M}}^2 = \frac{C}{\gamma} \, \frac{( p - 1 )}{ ( C - 1 )}
\left[ 1 + \frac{ b ( C^2 - 1 )}{( p - 1 )} \right]
=
\frac{5 C}{3} \, \frac{( p - 1 )}{ ( C - 1 )}
\left[ 1 + \frac{ b ( C^2 - 1 )}{( p - 1 )} \right]
\, .
\label{eq:magmach}
\end{equation}
As before, final expressions apply to the usual case of $\gamma = 5 / 3$.

\section{C.\quad Post-Shock Coulomb Heating of Electrons}
\label{sec:equi}

Assume that the electrons are not heated completely effectively at the shock, and immediately behind the shock
they have a temperature $T_e = T_0$, and that this value has been constant for the time it has taken the shock
to move through the region where the electrons are being heated.
Assume that all of the ions have the same temperature $T_{\rm ion}$.
Define the mean temperature of the ions and electrons as
${\bar{T}} \equiv \mu m_p P / (k \rho )$, where $P$ and $\rho$ are the total gas pressure and mass density, respectively, and $\mu m_p$ is the mean mass per particle.
Then, ${\bar{T}} = ( n_{\rm ion} T_{\rm ion} + n_e T_e ) / ( n_{\rm ion} + n_e )$.
Assume that ${\bar{T}} = $ constant in the region of interest immediate behind the shock.
Define $\tau \equiv T_e / {\bar{T}}$.
Then, the equation for the heating of the electrons by Coulomb collisions is
\citep[][equation~28]{WS09}
\begin{equation}
\frac{d \, \tau}{d \, t} = \frac{\tau^{-3/2} \left( 1 - \tau \right) }{ {\bar{t}}_{\rm eq} }
\, \frac{n_e}{n_{e2}}
\, ,
\label{eq:taueq}
\end{equation}
where ${\bar{t}}_{\rm eq}$ is the equipartition time (eq.~\ref{eq:teq}) evaluated at the mean
post-shock temperature ${\bar{T}}$ and with the proton density replaced by the total number density immediately behind the shock.

In our surface brightness model, the post-shock density is a power-law function of the radius,
$ n_e / n_{e2} = ( r / r_s )^{- p_i}$
(eq.~\ref{eq:emiss}).
In a narrow region behind the shock, we can reasonably assume that the post-shock velocity $v_2$ is a constant.
Then, $ d r = v_2 dt$.
The differential equation for the electron temperature is then
\begin{equation}
\frac{d \, \tau}{d \, (r / r_s )} = A \tau^{-3/2} \left( 1 - \tau \right) \left( \frac{r}{r_s} \right)^{-p_i}
\, ,
\label{eq:taudeq}
\end{equation}
where $ A \equiv r_s / ( v_2 {\bar{t}}_{\rm eq} )$ is a constant which effectively gives the timescale of the post-shock flow divided by the equipartition timescale.

The initial condition is that $\tau = \tau_0 \equiv T_0 / {\bar{T}} $ at $r = r_s$.
The solution is
\begin{equation}
f( \tau ) - f( \tau_0 ) = A
\left\{
\begin{array}{cc}
\frac{1}{p_i - 1} \left[ \left( \frac{r}{r_s} \right)^{1 - p_i} - 1 \right] & p_i \ne 1 \\
\ln \left( \frac{r_s}{r} \right) & p_i = 1
\end{array}
\right.
\, ,
\label{eq:tausol}
\end{equation}
where
\begin{equation}
f( \tau ) \equiv \ln \left| \frac{1 + \tau^{1/2}}{1 - \tau^{1/2}} \right|
-2 \tau^{1/2} - \frac{2}{3} \tau^{3/2}
\, .
\label{eq:eqfunc}
\end{equation}
While there is no obvious explicit solution for $\tau$ as a function of $r$, it is easy to solve for
$r$ as a function of $\tau$, which is just as useful:
\begin{equation}
\frac{r}{r_s} = 
\left\{
\begin{array}{cc}
\left\{ 1 + \frac{ \left( p_i - 1 \right) \left[ f( \tau ) - f( \tau_0 ) \right]}{A} \right\}^{1/ ( 1 - p_i )} & p_i \ne 1 \\
\exp \left[ - \frac{f( \tau ) - f( \tau_0 )}{A} \right] & p_i = 1
\end{array}
\right.
\, .
\label{eq:tausol2}
\end{equation}




\end{document}